\documentclass[prx,aps,twocolumn,10pt,superscriptaddress, nofootinbib, floatfix, a4paper, groupedaddress]{revtex4-2}
\pdfoutput=1
\usepackage[utf8]{inputenc}
\usepackage[T1]{fontenc}
\usepackage[english]{babel}
\usepackage[dvipsnames]{xcolor}
\usepackage{graphicx}
\usepackage{bm}
\usepackage{amsmath}
\usepackage{amssymb}
\usepackage{amsfonts}
\usepackage{amsthm}
\usepackage{bbold}
\usepackage[ruled,vlined]{algorithm2e}
\usepackage{color}
\usepackage{hyperref}
\usepackage{booktabs}
\usepackage{eqparbox}
\usepackage[mathlines]{lineno}
\usepackage{pgfplots}
\pgfplotsset{compat=newest} 
\usepgfplotslibrary{groupplots,dateplot} 
\usepackage{tikz}
\usetikzlibrary{patterns,shapes.arrows}
\usetikzlibrary{positioning}
\usepackage{dsfont}
\usepackage{tensor}
\usepackage{braket}
\usepackage{cleveref}
\usepackage{cases}
\usepackage{mathrsfs} 
\usepackage[scr=boondoxo,scrscaled=1.05]{mathalfa} 
\usepackage{enumitem}
\usepackage{placeins}
\usepackage{nicefrac}
\usepackage{upgreek}
\usepackage{microtype}
\usepackage{textcomp}
\usepackage{ragged2e}
\usepackage[marginal,norule]{footmisc}
\usepackage{xkvltxp}
\usepackage{quoting}
\usepackage{csquotes}
\usepackage{dcolumn}
\usepackage[normalem]{ulem}
\usepackage{calculator}
\usepackage{siunitx}
\usepackage{orcidlink}
\DeclareSIUnit\eVperc{\eV\per\clight}
\DeclareSIUnit\clight{\text{\ensuremath{c}}}
\hypersetup{
  linktoc = all,
  colorlinks = true,
  allcolors = RoyalPurple
 }
\makeatletter\let\@fnsymbol\@arabic\makeatother
\DeclareUnicodeCharacter{2212}{−}
\graphicspath{{./Images/}{./Plots/}}
\newtheorem{theorem}{Theorem}

\newcommand{\norm}[1]{\left\lVert#1\right\rVert_\mathrm{op}}
\newcommand{\normsmall}[1]{\lVert#1\rVert_\mathrm{op}}

\DeclareFontFamily{U}{mathc}{}
\DeclareFontShape{U}{mathc}{m}{it}%
{<->s*[1.03] mathc10}{}
\DeclareMathAlphabet{\mathcal}{U}{mathc}{m}{it}

\DeclareMathOperator{\Com}{\mathcal{C\mkern-3mu om}}

\begin{document}

\title{Hierarchical divide and conquer quantum approach to combinatorial optimization problems with tunable reduction}

\author{Mathias Schmid \orcidlink{0009-0000-7004-1914}}
\email{mathias.schmid@fau.de}
\affiliation{Physics Department, Friedrich-Alexander-Universität Erlangen Nürnberg, Germany}

\author{Naeimeh Mohseni \orcidlink{0000-0003-3373-4572}}
\thanks{present address: E.ON Digital Technology GmbH}
\affiliation{Physics Department, Friedrich-Alexander-Universität Erlangen Nürnberg, Germany}

\author{Michael J. Hartmann \orcidlink{0000-0002-8207-3806}}
 \email{michael.j.hartmann@fau.de}
\affiliation{Physics Department, Friedrich-Alexander-Universität Erlangen Nürnberg, Germany}

\begin{abstract}
Combinatorial optimization is 
considered a promising class of problems in which quantum computers can show significant advantages. However, problems of practical relevance typically have more variables than current or foreseeable quantum computers have qubits. Here we introduce a divide and conquer approach that partitions the optimization problem into subgraphs that can be represented on smaller quantum processors.
We then find all states of the subgraphs that can possibly be part of the solution to the entire problem by determining the cost or energy ranges in which the local subgraph energies of these states must be contained. 
This allows us to reduce the problem by only considering the subspace spanned by these states.
We then recombine the system using a binary encoding for each subgraph with a local energy ordering.
This process can be iterated until no further reduction is possible. We also find that the number of necessary qubits can be reduced further when only retaining states in a fraction of the relevant energy range at very little expense in terms of approximation ratio to the global ground state.
In numerical simulations, we 
find that our approach allows us to solve combinatorial optimization problems on weighted random 3-regular graphs with $|\mathcal{V}|=40$ discrete variables on $\sim |\mathcal{V}| / 4$ qubits while retaining a possible approximation ratio of $\sim99.9\%$.  We also observe an increasing reduction with larger system sizes.
\end{abstract}

\maketitle

\section{Introduction}\label{sec:intro}
Combinatorial optimization problems are often computationally hard but appear in many areas of science, logistics, and industry, including planning \cite{zgurovsky2018combinatorial} and internet data packet routing \cite{peis2009packet}. 
Quantum algorithms currently explored to solve such problems include quantum phase estimation \cite{ibm2018}, the adiabatic quantum algorithm \cite{Farhi2001}, and variational quantum algorithms, such as the Quantum Approximate Optimization Algorithm (QAOA) \cite{QAOA} or the Filtering Variational Quantum Eigensolver (F-VQE) \cite{filtering_case_study}.

\begin{figure*}[t!]
\centering
\includegraphics[scale=0.043]{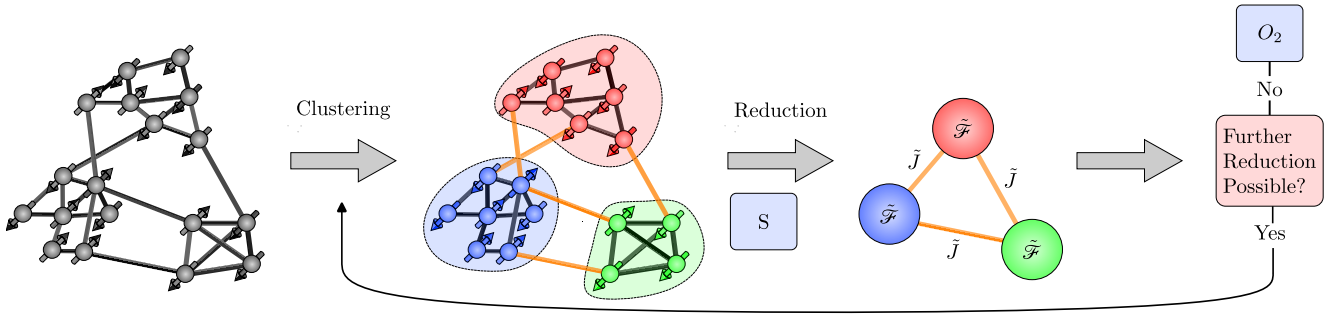}
\caption{Schematic representation of our approach. The combinatorial problem instance is represented as a graph. Using a community detection algorithm (\cref{sec:clustering}) this graph is divided into subgraphs, see step ''Clustering'' in the illustration. In a subroutine S (\cref{fig:subroutine}) the lowest-lying energy states within a predetermined energy range are determined and then represented by a new reduced System $\tilde{\mathscr{F}}$, see step ''Reduction''. If no further reduction is possible the whole system is solved using an optimizer $O_2$, otherwise it is again split into subgraphs and the process is iterated until no further reduction is possible.}
\label{fig:Ansatz1}
\end{figure*}

For solving a combinatorial optimization problem with quantum computers, one typically represents the cost function, which is evaluated on a finite solution space, namely size $n$ bit strings, as the expectation value of a Hamiltonian acting on $n$ qubits that replace the classical bits. Hence, the optimal solution to the problem is represented by the lowest energy eigenstate of the representing Hamiltonian \cite{Boolean_Hamilton}.

However, the number of variables in many instances of practical relevance vastly exceeds the number of qubits that current or foreseeable quantum computers can offer. For industrial applications, combinatorial optimization problems of interest often contain tens of thousands of variables.
It is thus of great importance to develop methods for running quantum algorithms on devices that contain much fewer qubits than the original algorithm would require.
There are multiple directions that can be taken to achieve this goal. 
 Since classical solvers are heavily relying on pruning the search space, there has been an increase in the sentiment that quantum approaches could benefit from similar approaches, including hybridization \cite{Smith-Miles_2025}. Alternatively, the solution space can be compressed using a Pauli-correlation encoding \cite{Pauli-Encoding} at the expense of increased sample complexity. 

Here we use ideas from community detection \cite{Louvain} to introduce a "divide and conquer" (DC) method for solving combinatorial optimization problems whose number of variables exceeds the number of available qubits on the quantum device by reducing the configuration space in a controlled and tuneable way. 
Preceding explorations of DC approaches in hybrid quantum algorithms and simulated annealing can be classified by the problem classes they aim to solve, how they divide the problem into smaller ones, how they conquer the subproblems, how they combine the subproblems to obtain a global solution, and whether the procedure can be performed iteratively  \cite{saleem2021quantum,zintchenko2015,DeepVQE,DC_QAOA_qsr_sampling_sub_disto, zhou2023qaoa, jo2022divide, guerreschi2021solving}. The types of problem most frequently studied are quadratic unconstrained binary optimization (QUBO) \cite{guerreschi2021solving, zintchenko2015, jo2022divide} and purely quadratic Hamiltonians that are equivalent to weighted MaxCut \cite{max_cut_quadratic,DC_QAOA_qsr_sampling_sub_disto, zhou2023qaoa}, which can both be represented on graphs. 

In the division step, the dominant idea is to consider disjunct subsets of nodes that are connected with weak couplings to different subsets. However, combinatorial optimization problems are often frustrated and the ground state of the full system is not necessarily a direct product of states that minimize the energies of the subsystems. Existing approaches can either not cope with frustration or offer methods that scale exponentially in the number of boundary qubits in the subsets. In one example, the $\mathds{Z}_2$ symmetry of MaxCut is used to reduce the local space to two states \cite{zhou2023qaoa} after solving the subset problem. In  \cite{DeepVQE} a larger local subspace is produced by applying local excitations on the boundary of the found local ground state. Yet, due to these limitations, such approaches do not fully cover the frustration of the system. On the other hand, the approach in \cite{guerreschi2021solving} divides the subsystems into boundary and bulk variables, before performing local optimizations of the bulk for all boundary configurations. In the end, an optimization over all boundary variables is used to find a global solution. This yields a constant variable reduction, but it requires solving a number of local optimizations that grows exponentially in the number of qubits on the boundary.

In contrast, the DC approach, which we introduce here, guarantees that  the global optimum is retained in all reduction steps, even for highly frustrated problems. At the same time, it allows us to tune the targeted qubit reduction and hence enables treatments of highly frustrated problems with limited resources. We find in multiple examples that this new feature does not significantly affect the quality of the solution, for a sizable tuning range. Additionally, our approach can also be used for the large class of polynomial unconstrained binary optimization (PUBO) problems, which are representable by hypergraphs.

We divide the system into subsystems, using heuristic community detection algorithms, that result in disjunct graphs/hypergraphs with strong intra-community connections and weak couplings across communities. For problems on graphs, on which we focus in our explanation of the method, we use the so called Louvain algorithm \cite{Louvain}. 
After the division of the system, we introduce an energy cut-off that depends on the operator norm of an interaction Hamiltonian for each considered subsystem. Finding all local states within an energy range given by this cut-off point with a (quantum) optimizer $O_1$ guarantees that the global ground state is contained in this set, while all states with higher local energy can be discarded. We find that due to the $\mathds{Z}_2$ symmetry in purely quadratic Hamiltonians the cut-off is reduced by a factor $2$.  
After finding all relevant states in the subsystems we map them onto a new reduced qubit basis using a binary encoding with local energy ordering. 

In this reduced basis, we build a new optimization problem by considering the action of the original Hamiltonian in this new basis. This new optimization problem could then be solved directly by a (quantum) optimizer $O_2$, which can be the same as $O_1$, or different. The distinction is made to respect the difference in problem structure between the original problem and the reduced version. We, however, propose a hierarchical approach, where each local system is contracted to a single vertex and another round of community detection is performed to find a new (reduced) set of communities, for which the low energy subspaces can be solved with optimizer $O_2$.  This procedure is iterated until no further reduction is possible, at which point the global solution is searched. 

\begin{figure*}[t]
\centering
\includegraphics[scale=0.05]{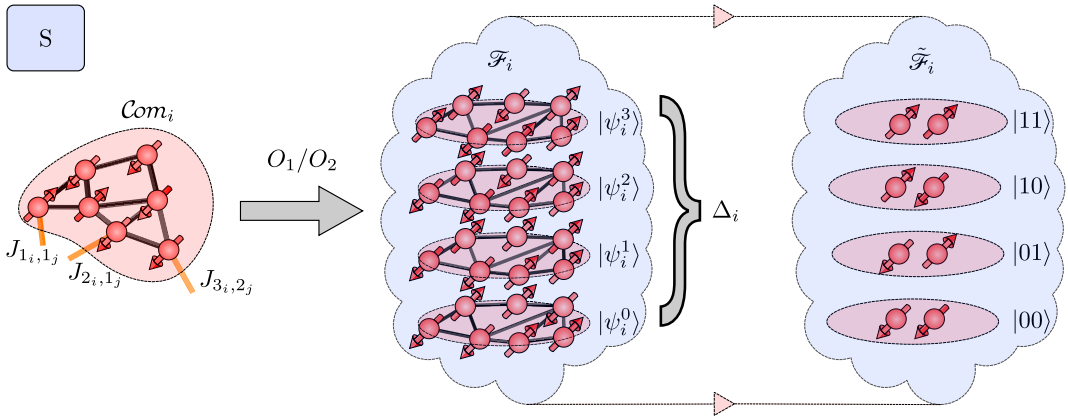}
\caption{The local subroutine S of our algorithm. For a given Community $\Com_i$ we determine the cut-off range (Here for a quadratic Hamiltonian $\Delta_i=|J_{1_i,1_j}|+|J_{2_i,1_j}|+|J_{3_i,2_j}|$) and determine all low-lying states within ($\mathscr{F}_i$) using an optimizer $O_1$. Then we represent the local solutions as computational basis states using a local energy ordering with fewer qubits $\tilde{\mathscr{F}_i}$. If it is not the first iteration, a different optimizer $O_2$ might be used, see \cref{subsec:iteration}.}
\label{fig:subroutine}
\end{figure*}

Further, we find that reducing the cut-off range by a constant factor retains a very good approximation ratio up to a certain point, resulting in a novel and canonical way to exchange search space size and qubit number for approximation ratio with a sweet spot at half of the cut-off range for the systems considered. A limiting case of our approach where only local degenerate ground states are retained in a MaxCut problem has been explored in \cite{zhou2023qaoa}. A sketch of the approach is provided in \cref{fig:Ansatz1} with the local subroutine depicted in \cref{fig:subroutine}.

In numerical experiments, we find that our approach allows us to solve combinatorial optimization problems with $|\mathcal{V}|$ discrete variables (weighted random 3-regular graphs of $|\mathcal{V}|$ vertices) on $\sim |\mathcal{V}| / 4$ qubits for problems with $|\mathcal{V}| \gtrsim 40$ while retaining a possible approximation ratio of $\sim99.9\%$. Our results indicate that even larger reductions can be expected for larger problem instances. This behavior makes our approach a promising strategy for solving optimization problems closer to practical relevance on much smaller quantum processors than a direct representation would require.

\section{Approach} \label{sec:approach}
We consider combinatorial optimization problems $(\mathcal{X}=\{0,1\}^n, \mathcal{C})$ on bit strings $z \in \mathcal{X}$ with a cost function $\mathcal{C}(z)$, for which the quantum version can generally be represented by a Hamiltonian \cite{Boolean_Hamilton} acting on qubits of the form 
\begin{equation}\label{eq:Pubo_hamiltonian_main}
    H=\sum_{S\subseteq [n]}J(S)\prod_{j\in S}Z_j,
\end{equation}
where $J(S)\in \mathds{R}$, $[n]=\{1,\dots,n\}$ and $Z_j$ denotes a Pauli-Z operator on qubit $j$, see \cref{sec:app-graphs}.
Yet, for illustrative purposes and because it corresponds to the famous maximum cut problem \cite{max_cut_quadratic}, we first outline our approach for Hamiltonians of the form
\begin{equation}\label{eq:two_body_ham2}
    H=\sum_{a<b}J_{a,b}\, Z_a\, Z_b
\end{equation}
and discuss the extension to general cost function Hamiltonians in \cref{sec:extensions}.

In our approach, we first represent these problem instances on graphs, see \cref{fig:Ansatz1} and \cref{sec:app-graphs} for details, and use the Louvain clustering algorithm \cite{Louvain}, which is based on modularity optimization, to divide the graph into disjoint subgraphs, see also \cref{sec:clustering}.
After clustering the system into subgraphs, we can split the problem Hamiltonian accordingly into parts $H_i^{(C)}$, acting on the $i$-th community(subgraph) denoted by $\Com_i$, and terms $H_{i,j}^{(I)}$, describing the interactions between communities, 
\begin{equation}\label{eq:decompose_into_coms}
 H=\sum_{i=1}^{N} H_i^{(C)} +\sum_{i<j}^{N} H_{i,j}^{(I)},
\end{equation}
where the indices $i$ and $j$ label the communities, $N$ is the number of communities, and 
\begin{align}
H_{i}^{(C)} = & \sum_{n_i=1}^{N_{i}}\sum_{m_i=1}^{N_{i}}J_{n_i,m_i}Z_{n_i}Z_{m_i},\\
 H_{i,j}^{(I)} = & \sum_{n_i=1}^{N_{i}}\sum_{m_j=1}^{M_{j}}J_{n_i,m_j}Z_{n_i}Z_{m_j}. \label{eq:interaction_ham}
\end{align}
Here, $J_{n_i,m_j}$ is the coupling of the $n$-th qubit in $\Com_i$ with the $m$-th qubit in $\Com_j$. The total number of qubits in community $i$ and $j$ is $N_i$ and $M_j$, respectively. Note that we have not explicitly written direct products with identity operators acting on the remaining communities.

Since the Hamiltonian $H$ in Eq. \eqref{eq:two_body_ham2} contains only Pauli-Z operators, its ground state $\ket{\psi_0}$ is a computational basis state and in particular a direct product of eigenstates $\ket{\psi_{i}^{\mu}}$, of the individual parts $H_i^{(C)}$,
\begin{equation} \label{eq:ground-state-as-product}
  \ket{\psi_0} \in \bigotimes_{i=1}^N \{\ket{\psi_{i}^{\mu}}\}_{\mu}, \,\,\,\text{with} \,\,\, H_i^{(C)}\ket{\psi_{i}^{\mu}} = E^{\mu}_i \ket{\psi_{i}^{\mu}},
\end{equation}
where $E_i^{\mu}$ are the energy eigenvalues of $H_i^{(C)}$. We use here a notation where $E_{i}^0 \leq E_{i}^1\leq E_{i}^2 \leq \dots$, and hence $\ket{\psi_{i}^{0}}$ is the local ground state of $H_i^{(C)}$. 
Importantly, due to frustration, which typically occurs in hard problems, the overall ground state $\ket{\psi_0}$ is not a direct product of the local ground states, $\ket{\psi_0} \not= \bigotimes_i \ket{\psi_{i}^{0}}$.

\subsection{Hilbert space reduction}\label{subsec:Hilber_space_reduction}
To overcome the limitations of previous approaches, we generalize divide and conquer strategies to find all local eigenstates $\ket{\psi_{i}^{\mu}}$ that can possibly be part of the product ground state $\ket{\psi_0}$ in Eq. \eqref{eq:ground-state-as-product}. We determine these states from the following consideration and resulting theorem.

We can rewrite the Hamiltonian $H$ in Eq. \eqref{eq:decompose_into_coms} so that we focus our attention on a single community $\Com_i$ and treat the remaining communities as an environment,
\begin{equation}\label{eq:decomp_environment}
    H = H_i^{(C)}+ H_i^{(I)} + H_\mathscr{E},
\end{equation}
where the total interaction of $\Com_i$ with the remaining communities is given by
\begin{equation}
    H_i^{(I)} =  \sum_{j:j\neq i} H_{i,j}^{(I)},
\end{equation}
and all terms that do not act on qubits in $\Com_i$ form the environment,
\begin{align}
     H_\mathscr{E} &= H - H_i^{(C)}+ H_i^{(I)}.
\end{align}
Then the following theorem holds.
\begin{theorem}\label{thm:cut_off_two_body}
Given a two body Hamiltonian $H$ as in Eq. \eqref{eq:two_body_ham2} and its decomposition into $\Com_i$ and the environment as in Eq. \eqref{eq:decomp_environment}, a local state $\ket{\psi_i^\mu}$ with $H_i^{(C)}\ket{\psi_{i}^{\mu}} = E^{\mu}_i \ket{\psi_{i}^{\mu}}$, can only be part of the global ground state, if its local energy satisfies $E_i^\mu\in [E_i^0, E_i^0+\Delta_i]$, where the spectral range $\Delta_i$ is given by
\begin{equation}
        \Delta_i= \norm{H_i^{(I)}},
\end{equation}
with $\norm{\cdot}$ taken as the operator norm.
\end{theorem}

The proof can be found in \cref{sec:cut_off_proof}, and an extension of this theorem to general cost function Hamiltonians is given in \cref{sec:extensions}.

As a consequence, by finding all local eigenstates $\ket{\psi_{i}^{\mu}}$ with energy $E_{i}^\mu\in[E_{i}^0,E_{i}^0+\Delta_i]$, we can ensure that all relevant states that can possibly be part of the global optimum are in the set of found states. Hence, the sought ground state of the Hamiltonian $H$ in Eq. \eqref{eq:two_body_ham2} is also in this set,
\begin{equation}
    \ket{\psi_0} \in\bigotimes_{i=1}^N \mathscr{F}_i, 
\end{equation}
where $\mathscr{F}_i$ is the local reduced Hilbert space spanned by the set of computational basis states $\ket{\psi_{i}^{\mu}}$ with $E_{i}^\mu \in [E_{i}^0,E_{i}^0+\Delta_i]$. Therefore, our strategy is to retain all $\ket{\psi_{i}^{\mu}} \in \mathscr{F}_i$ and discard states with higher energy to reduce the problem.

Note that $H_i^{(I)}$ is representable as a bipartite graph and determining $\normsmall{H_i^{(I)}}$ for bipartite graphs with arbitrary edge weights is NP-hard \cite{bipartite_nphard}. Thus, in the following and in the simulations of \cref{sec:simulations}, we further upper bound the local cut-off ranges given by theorem \ref{thm:cut_off_two_body}.
Since Pauli Z operators have eigenvalues $\pm 1$, the operator norm of their product on different sites is given by $\norm{Z_iZ_j}=1$. Using triangle inequality and absolute homogeneity, the norm of the interaction Hamiltonian can thus be bounded as
\begin{equation}
    \norm{H_i^{(I)}}\leq \sum_{j:j\neq i}^N  \sum_{n_i=1}^{N_{i}}\sum_{m_j=1}^{M_{j}}|J_{n_i,m_j}| .
\end{equation} 
Another outlet to reduce the qubit count and runtime is a reduction of the cut-off range 
\begin{align} \label{eq:eta-def}
    \Delta_i\rightarrow \eta \Delta_i \: \:  \text{with} \: \: 0\leq\eta\leq1 .
\end{align}
This can lead to a possible exclusion of the exact ground state, but as seen in \cref{sec:simulations}, it still produces very good approximations in the cases considered while significantly reducing the qubit count. Further analysis of the reasons for the very good approximation quality of $\eta <1 $ is discussed in \cref{sec:comments_on_eta}. Note that in the limit of $\eta = 0$ and only considering two local ground states one would recover the method of \cite{zhou2023qaoa}. 

\subsubsection{Building \texorpdfstring{$\mathscr{F_i}$}{F}}
To find all $\ket{\psi_{i}^{\mu}} \in \mathscr{F}_i$ for each subsystem with problem Hamiltonian $H_i^{(C)}$ we use an iterative procedure, where we first find the ground state $\ket{\psi_{i}^{0}}$ and then all excited states $\ket{\psi_{i}^{\mu}}  \in \mathscr{F}_i$ successively by adding energy penalties for the already found states to the problem Hamiltonian.
For first finding the ground state, we use an optimizer or algorithm $O_1$ to determine it. In the proof of principle numerical experiments, see \cref{sec:simulations}, we chose the Filtering Variational Quantum Eigensolver (F-VQE)\cite{filtering} for this. However, most optimizers for quadratic Hamiltonians are suitable, including the adiabatic algorithm \cite{Farhi2001} or QAOA \cite{QAOA}. For finding the remaining states, we denote the set of found states as $\mathscr{F_i^\prime}$ and choose the Hamiltonian for the next iteration to be
\begin{equation}H_i^{(C,\mathscr{F_i^\prime})}=H_i^{(C)}+\sum_{\ket{\psi_{i}^{\mu}} \in \mathscr{F_i^\prime}} p_i^{\mu} \ket{\psi_{i}^{\mu}}\bra{\psi_{i}^{\mu}},
\end{equation}
where $p_i^{\mu}$ are energy penalties for which we explain our choice below.  
We iterate this procedure until all states with $E_{i}^0 \leq E^{\mu}_i \leq E_{i}^0+\Delta_i$ are found. The fact that all eigenstates of $H_i^{(C)}$ are computational basis states makes this computation efficient. 
Every measured bit string during an optimization run with $E_{i}^0 \leq E^{\mu}_i \leq E_{0}^i+\Delta_i$ corresponds to one of the sought states and is included in $\mathscr{F_i^\prime}$. This strategy can be employed not only for variational algorithms \cite{filtering,QAOA}, but also for the adiabatic algorithm  \cite{Farhi2001}, where slightly non-adiabatic sweeps can be used to sample low lying excited states.
In practice, $E^0_i$ is taken to be the lowest measured energy from $\mathscr{F_i^\prime}$ and the iterations are continued until no more bit strings with energy $\leq E_{0_i}+\Delta_i$ are measured during an optimization run at which point we set $\mathscr{F_i \equiv F_i^\prime}$.
 
When choosing the magnitude of the penalty term $p^{\mu}_i$ one should consider two things. On the one hand, it has to be large enough so that the state is not found twice, and on the other hand, choosing the penalty too large changes the energy landscape and negatively impacts the optimization process. The first consideration can be quantified by the lower bound
$p_{i}^\mu>E_{i}^0-E_{i}^\mu+\Delta_i$, which can be calculated after obtaining $\ket{\psi_i^\mu}$. In practice, we heuristically chose $p_i^\mu=\Delta_i +c_1|E^\mu_i|+c_2$ with small constants $c_1, c_2 \ll 1$. Also note that the penalties are included when evaluating the measured computational basis states, which is possible because the Hamiltonian is diagonal.

Furthermore, since we are interested not only in the ground state of $H^{(C,\mathscr{F_i^\prime})}_i$ but all lower-lying states in each iteration, our iterative procedure provides some robustness against cases where the optimizer is unable to find the lowest possible state. This robustness emerges since there is another chance to find that state in the next iteration with a changed energy landscape that includes penalties for states which were found instead. 

\subsection{Building an optimization problem in the truncated Hilbert space}\label{subsec:reduced_hamiltonian}
\begin{figure}[t!]
\centering
\includegraphics[scale=0.052]{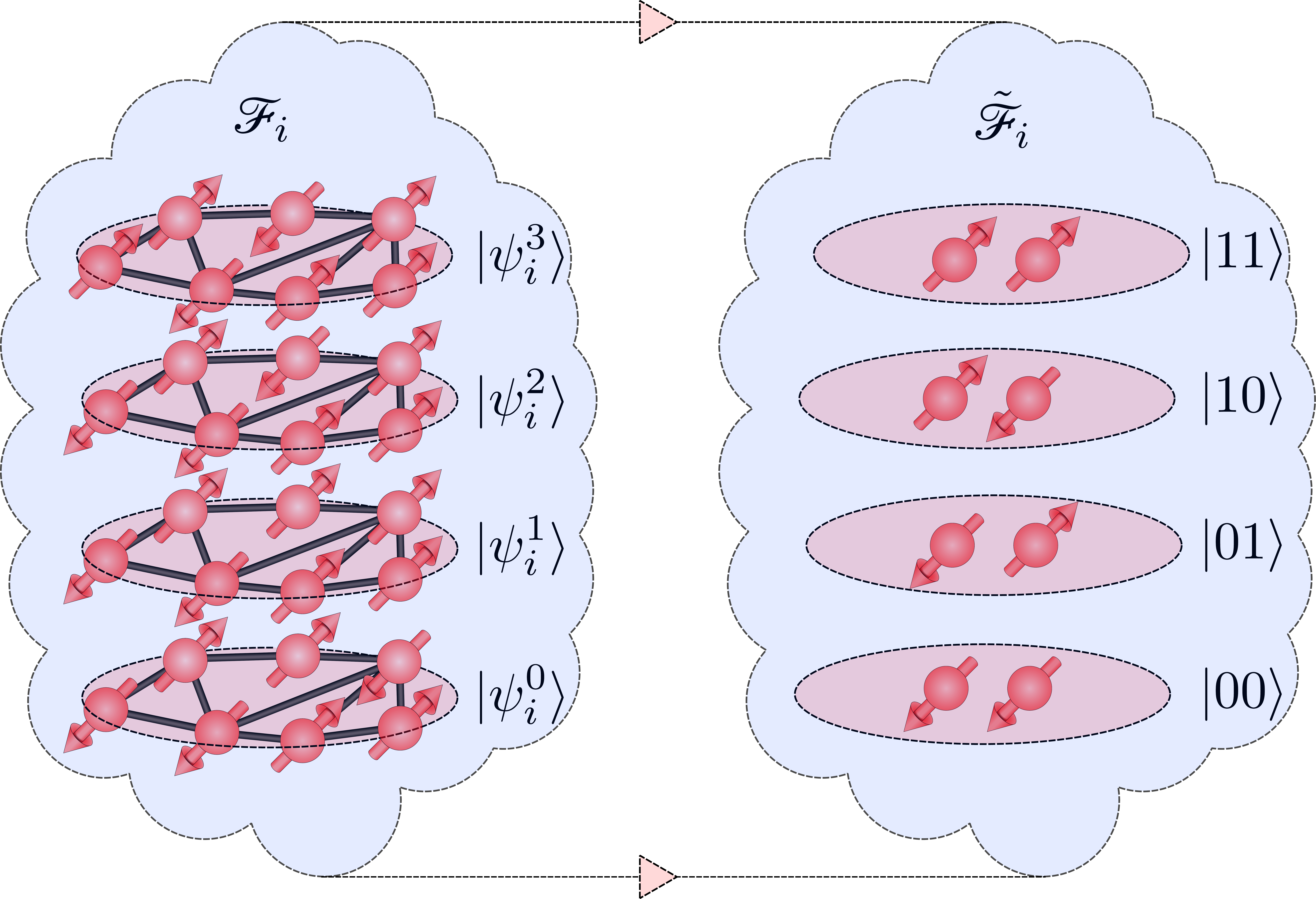}
\caption{The low lying set of eigenstates $\ket{\psi_i^\mu} \in \mathscr{F}_i$ of a community $\Com_i$ (left) is cast into the computational basis of a reduced number of qubits $\tilde{\mathscr{F}}_i$ (right) by using a binary encoding with a local energy ordering.}
\label{fig:storage}
\end{figure}

We now discuss how the action of the original problem Hamiltonian $H$ in Eq.\eqref{eq:two_body_ham2} on the reduced Hilbert space can be cast into a new, smaller optimization problem with reduced Hamiltonian $H_\mathrm{red}$. 
Since we want to represent the Hamiltonian $H_\mathrm{red}$ on as few qubits as possible, we associate the states found $\ket{\psi_{i}^{\mu}}$ with the computational basis states of a reduced number of qubits  $\tilde{M}_i$, which must be sufficient to accommodate the reduced Hilbert space $\mathscr{F}_i$ for each community. Denoting the dimension of $\mathscr{F}_i$ by $d_i = \text{dim}(\mathscr{F}_i)$, one finds
 \begin{equation}
     \tilde{M}_i =\lceil\log_2(d_i)\rceil
\end{equation}
where $\lceil x \rceil$ denotes the smallest integer that is larger than $x$ and we set $\tilde{M_i}=1$ if $d_i=1$. We choose the following identification for these states,
\begin{equation}\label{eq:binary_representation}\ket{z_{i,1},z_{i,2},\dots,z_{i,\tilde{M}_i}} \hat{=} \ket{\psi_{i}^{\mu}}
\end{equation}
for $z_{i,\alpha} \in \{0,1\}$ where
$z_{i,1} 2^{\tilde{M}_i-1} + z_{i,2} 2^{\tilde{M}_i-2} + \dots + z_{i,\tilde{M}_i} = \mu$ is the binary representation of the state index $\mu$, i.e. a binary encoding with local energy ordering.
If the number of energy levels found is not a power of two, one can either introduce cost energies that fill up the remaining spots,
which ensures that non-valid states are heavily penalized, or repeat the lowest lying states s.t. $(z_{i,1} 2^{\tilde{M}_i-1} + z_{i,2} 2^{\tilde{M}_i-2} + \dots + z_{i,\tilde{M}_i} ) \mod d_i = \mu$.
We denote this qubit representation of the local reduced Hilbert space as $\tilde{\mathscr{F}_i}$ and $\dim(\tilde{\mathscr{F}_i})=\tilde{d}_i$, a visualization is provided in \cref{fig:storage}.

The action of $H$ as in Eq. \eqref{eq:two_body_ham2} on $\mathcal{H}_\mathrm{red} = \bigotimes_{i=1}^N \tilde{\mathscr{F}}_i$ can then be written as
\begin{equation}\label{eq:reduced_hamiltonian}
    H_\mathrm{red} = \sum_{i=1}^N \tilde{H}_i^{(C)}   + \sum_{i<j}^{N} \tilde{H}_{i,j}^{(I)} ,
\end{equation}
where all matrices in Eq. \eqref{eq:reduced_hamiltonian} are diagonal. The diagonal elements of the reduced community Hamiltonian $\tilde{H}_i^{(C)}$ are the corresponding eigenenergies 
\begin{equation}\label{eq:reduced_community_hamiltonian}
    \tilde{H}_i^{(C)} \ket{\psi^\mu_i} =  E^{\mu}_i \ket{\psi^\mu_i}
\end{equation}
 and the reduced interaction reads
\begin{equation}\label{eq:coupling-matrix}
    \tilde{H}_{i,j}^{(I)} \ket{\psi^\mu_i,\psi^\nu_j} =  \sum_{n_i=1}^{N_{i}}\sum_{m_j=1}^{M_{j}}J_{n_i,m_j}\chi_{n_i,m_j}^{\mu,\nu} \ket{\psi^\mu_i,\psi^\nu_j} ,
\end{equation}
where $
    \chi_{n_i,m_j}^{\mu,\nu} = \bra{\psi_{i}^{\mu}, \psi_{j}^{\nu}} Z_{n_i} Z_{m_j} \ket{\psi_{i}^{\mu},\psi_{j}^{\nu}} = \pm 1
$
is the sign of the coupling for the inter-community edge $(n_i,m_j)$ when the system is in the state $\ket{\psi_{i}^{\mu},\psi_{j}^{\nu}}$.

This reduced Hamiltonian $H_\mathrm{red}$ can now be used to generate the gates for some optimizer $O_2$ of the reduced problem. 
$H_\mathrm{red}$ can also be written in terms of Pauli-Z operators 
via the standard procedure to expand Hermitian matrices in a basis of strings of Pauli and identity matrices. Here, however, since the matrices of $H_\mathrm{red}$ are diagonal, direct products of Pauli-Z  and identity matrices form a sufficient basis. 

It is worth mentioning that the reduced Hamiltonian is no longer built of terms that are only quadratic in Pauli-Z operators but in general may include all products of Pauli-Z's on the active reduced qubits. 
This is expected since $H_\mathrm{red}$ was constructed by a local energy ordering and should be taken into account when choosing the optimizer $O_2$. Also note that the multidimensional array $\chi$ has \begin{equation}\label{eq:n_chi}
    N_\chi=\sum_{i<j}^{N} \tilde{d}_i\tilde{d}_jN_{i,j}^{\mathrm{Edges},\mathrm{(I)}} \leq N_\mathrm{total}^{\mathrm{Edges},\mathrm{(I)}} \max_i (\tilde{d}_i^2)
\end{equation}
elements. Here, $N_{i,j}^{\mathrm{Edges},\mathrm{(I)}}$ denotes the number of edges between communities $\Com_i$ and $\Com_j$.  $N_\mathrm{total}^{\mathrm{Edges},\mathrm{(I)}}$ is the total number of inter-community edges. If $O_2$ is chosen to be a brute force optimizer, the number of evaluations is given by $N_\mathrm{eval}=\prod_i \tilde{d}_i \gg N_\chi$. Furthermore, depending on the choice of $O_2$ it might not be necessary to calculate the entire array $\chi$, but only the entries that correspond to the solution candidates.

\subsection{Iteration}\label{subsec:iteration}
Instead of solving the reduced problem in one step, we can also choose to divide the problem again into a set of communities and find the lowest lying states for each. This leads to an iterative repetition of our strategy. 

To build the iteration, we again need to associate the problem with a graph. This can be achieved by considering all states $\ket{\psi_{i}^{\mu}} \in \tilde{\mathscr{F}}_i$ as a single $\tilde{d}_i$-level quantum system, located at one vertex. The weight of the edge connecting two such $\tilde{d}_i$-level and $\tilde{d}_j$-level vertices can then be chosen as the operator norm of the coupling matrix $\tilde{H}_{i,j}^{(I)}$ \footnote{If the method with penalty entries was chosen, these should of course be neglected.},
\begin{equation}\label{eq:iteration_clustering_coupling}
    \tilde{J}_{i,j} =\norm{ \tilde{H}_{i,j}^{(I)} }
\end{equation}
Based on these edge weights, one can again cluster the vertices into communities. If it was chosen not to calculate all entries in $\chi$, an upper bound for the operator norm in Eq. \eqref{eq:iteration_clustering_coupling} can be used by summing the absolute values of the corresponding edge weights.
\begin{equation}\label{eq:iteration_clustering_coupling_alt}
    \norm{ \tilde{H}_{i,j}^{(I)} } \leq \sum_{n_i=1}^{N_{i}}\sum_{m_j=1}^{M_{j}}|J_{n_i,m_j}|
\end{equation}
After the clustering process, one can then solve for all states with low enough energy in each cluster. 
While the weights in Eq. \eqref{eq:iteration_clustering_coupling} are used for the clustering process, solving for low energy states within the clusters is done via the local Hamiltonians of the clusters. For the new community $\Com_l$ this Hamiltonian reads
\begin{equation}\label{eq:iter_local_hamiltonian}
    H_{l}^{(C,\mathrm{it})}=\sum_{i\in \Com_l} \tilde{H}_i^{(C)}+\sum_{i,j\in \Com_l:\, i<j} \tilde{H}_{i,j}^{(I)} \, ,
\end{equation}
which is a local version of the Hamiltonian in Eq. \eqref{eq:reduced_hamiltonian}. Note that since each vertex represents a reduced community with space $\tilde{\mathscr{F}}_i$, the number of qubits required to represent this local Hamiltonian is not the number of vertices within the community, but the sum of storage qubits $\sum_{i \in \mathcal{Com}_l} \tilde{M}_i$. Analogously, the interactions between $\mathcal{Com}_l$ and $\mathcal{Com}_k$ are given by
\begin{equation}
    H_{l,k}^{(I,\mathrm{it})}= \sum_{i\in \Com_l, j\in \Com_k} \tilde{H}_{i,j}^{(I)}
\end{equation}
The strength of the interaction Hamiltonian can again be upper bounded, as in the first step, by
\begin{equation}
    \norm{H_{l,k}^{(I,\mathrm{it})}}\leq  \sum_{i\in \mathcal{Com}_l, j\in \mathcal{Com}_k} \tilde{J}_{i,j} \, ,
\end{equation}
The new cut-off range for the local community Hamiltonian in Eq. \eqref{eq:iter_local_hamiltonian} thus is
\begin{equation}
     \Delta_l^\mathrm{it} =\sum_{k=1:k\neq l}^{N^\mathrm{it}}  \sum_{i\in \mathcal{Com}_l, j\in \mathcal{Com}_k} \tilde{J}_{i,j}.
\end{equation}
where $N^\mathrm{it}$ denotes the number of communities in the new iteration. This new cut-off follows directly from \cref{thm:cut_off_two_body}, where the symmetric properties are carried through the iterations, see \cref{sec:cut_off_proof}. 
After finding all the local states within $[E_{l}^{0,\mathrm{it}},E_{l}^{0,\mathrm{it}}+\Delta_l^\mathrm{it}]$ for all community Hamiltonians, we again build a reduced Hamiltonian
\begin{equation}\label{eq:reduced_ham_it}
    H_\mathrm{red}^\mathrm{it}=\sum_{l=1}^{N^\mathrm{it}} {\tilde{H}}_{l}^{(C,\mathrm{it})}+\sum_{l,k=1:\, l<k}^{N^\mathrm{it}} {\tilde{H}}_{l,k}^{(I,\mathrm{it})}
\end{equation}
Here, the reduced Hamiltonians ${\tilde{H}}_{l}^{(C,\mathrm{it})}$ and ${\tilde{H}}_{l,k}^{(I,\mathrm{it})}$ are built as described in \cref{subsec:reduced_hamiltonian}, where an identification of the states that were found in this iteration, with the original $\ket{\psi_l^{\kappa,\mathrm{it}}}\hat{=}\bigotimes_{i\in \mathcal{Com}_l}\ket{\psi_i^\mu}$ is required to calculate  $
    \chi_{n_l,m_k}^{\kappa,\eta} = \bra{\psi_{l}^{\kappa,\mathrm{it}},\psi_{k}^{\eta,\mathrm{it}}} Z_{n_l} Z_{m_k} \ket{\psi_{l}^{\kappa,\mathrm{it}},\psi_{k}^{\eta,\mathrm{it}}}.
$

The described process can be iterated, where high iteration numbers are suitable for larger graphs and graphs with hierarchical community structure. The number of iterations can either be fixed beforehand or determined by a recombination criterion. To promote qubit reduction, we chose to recombine only if:
\begin{enumerate}
    \item The recombined system would use fewer qubits than were needed at some prior point
    \item Another iteration would yield the same communities as the current iteration
    \item The clustering algorithm groups everything into one community
\end{enumerate}
The algorithm is summarized by the pseudo code in \cref{alg:approach}.

\begin{algorithm*}[t]
\SetAlgoLined
\caption{Divide and conquer for quadratic Hamiltonians}\label{alg:approach}
\KwIn{Problem instance $(\{0,1\}^n,\mathcal{C})$, Hyperparameter $\eta$}
\KwOut{Approximate GS solution $\tilde{\ket{\psi_0}}$ and its energy $\tilde{E_0}$}
\Begin{
Convert the problem into a graph\;
Use a clustering algorithm\;
Determine the cut-off energies $\Delta_i$ for each cluster\;
Solve the subsystems for all states with $E^\mu_i \in [E^0_i,E^0_i+\eta\Delta_i]$ using $O_1$\;
Represent the found states on a reduced number of qubits $\tilde{\mathscr{F}}_i$\;
Calculate the summands of the reduced Hamiltonian $H_\mathrm{red}$ (optional)\;
\eIf{Recombination criterion is met}{Solve the recombined system using $O_2$\;}{
\While{True}{
    Convert the reduced problem into a graph by contracting $\tilde{\mathscr{F}}_i$ into vertices and use $\tilde{J}_{i,j}$ as edges\;
    Use a clustering algorithm\;
    Determine the cut-off energies $ \Delta_i^\mathrm{it} $ for each cluster\;
    Solve the subsystems for all states with $E_{i}^{\mu,\mathrm{it}}\in[E_{i}^{0,\mathrm{it}},E_{i}^{0,\mathrm{it}}+\eta\Delta_i^\mathrm{it}]$  using $O_2$\;
    Represent the found states on a reduced number of qubits $\tilde{\mathscr{F}}_i$\;
    Calculate the summands of the reduced Hamiltonian $H_\mathrm{red}^\mathrm{it}$ (optional)\;
    \If{Recombination criterion is met}{\textbf{break}\;
}}
Solve the recombined system using $O_2$\;}
}
\end{algorithm*}

We would also like to mention that \cref{alg:approach} can be used in varied versions from its base form depending on the circumstances, e.g. instead of fixing the hyperparameter $\eta$ one can do a sweep starting from low values of $\eta$ until a saturation of the output energy is reached. It is also not strictly necessary to acquire every single state with a local energy value inside the given range to obtain a good approximation, e.g. one could do a pseudo adiabatic sweep \cite{Lucas_pseudo_adiabatic, pseudo_adiabatic} to obtain a good batch of low energy candidates, which are then used for further optimization. Furthermore, it is possible to adapt the algorithm to ensure compatibility with a device with fixed qubit number. Firstly, a community detection algorithm that takes into account community size limitations should be used \cite{resolution_cd,constraint_community_detection}. After the first iteration, one has to consider that every recombined system should also fit on the device. If this is not the case from the beginning, one would have to reduce $\eta$ for the subsystems in question or split the optimization of the recombined system into multiple parts. However, this scales exponentially with the number of surplus qubits.
Note that one could also simply use the base version of the algorithm as long as the community detection algorithm produces small enough communities and $\eta$ is chosen sufficiently small so that everything is executable on the quantum computer.

\section{Results} \label{sec:simulations}
\begin{figure*}[!ht]
\centering
\includegraphics[scale=0.35]{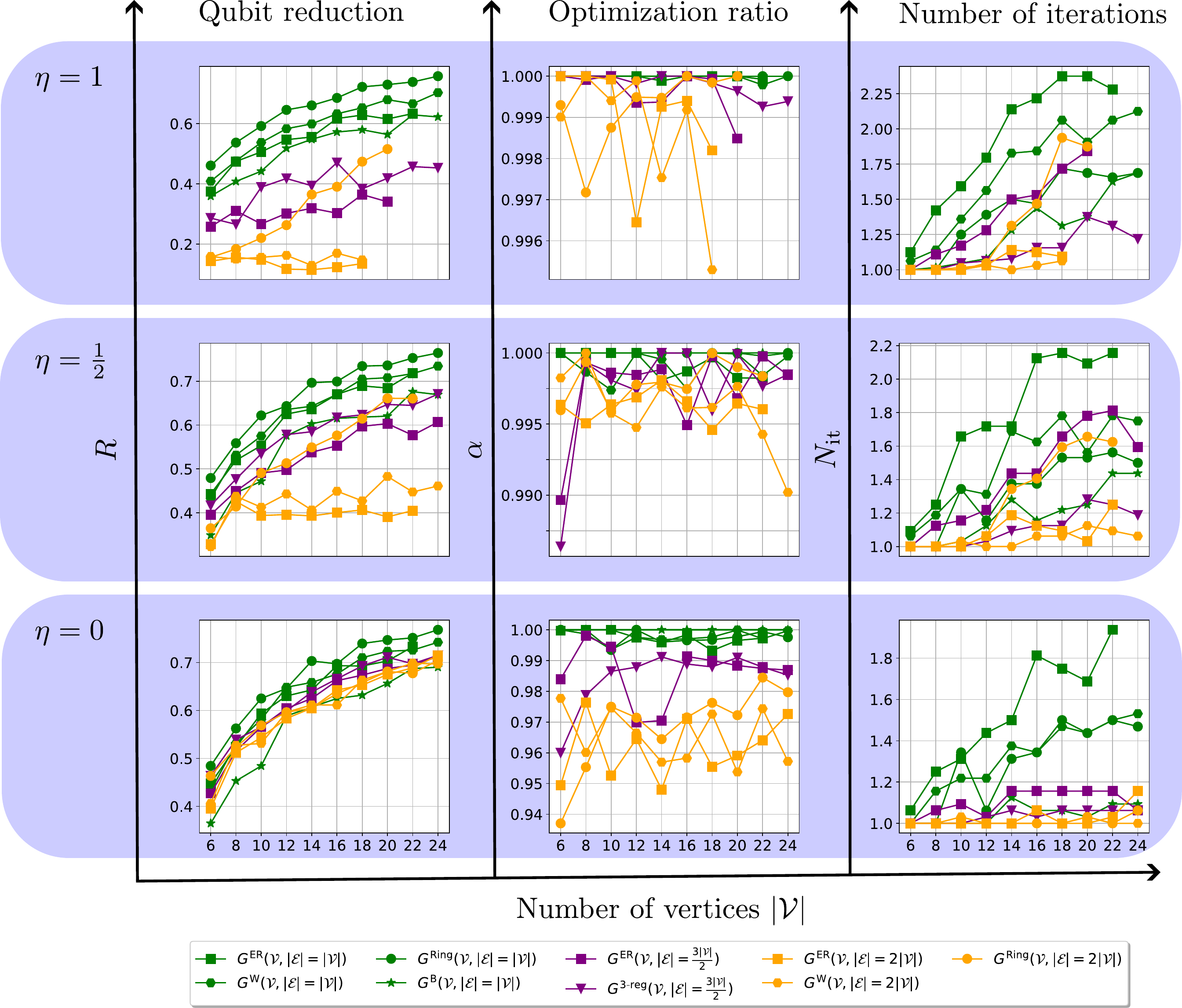}
\caption{Performance of the algorithm for $O_1=O_2=$F-VQE for a set of various weighted random graphs as indicated in the legend (see Sec. \ref{sec:random_graphs} for further explanations). Left column: qubit reduction $R$ as defined in Eq. \eqref{eq:qubit_red}. Middle column: approximation ratio $\alpha$ as defined in Eq. \eqref{eq:approx_ratio}. Right column: number of iterations in the algorithm $N_{\text{it}}$. Each row shows results for retaining different fractions $\eta$ of the energy ranges in the subsystems, where $\eta=0$ means that only the degenerate ground states are retained, where another optimization $O_2$ is still performed. Green lines correspond to graphs with underlying average degree 2, purple lines to graphs with degree 3 and orange lines to graphs with degree 4.
}
\label{fig:proof_of_principle}
\end{figure*}
\begin{figure*}[!ht]
\centering
\includegraphics[scale=0.35]{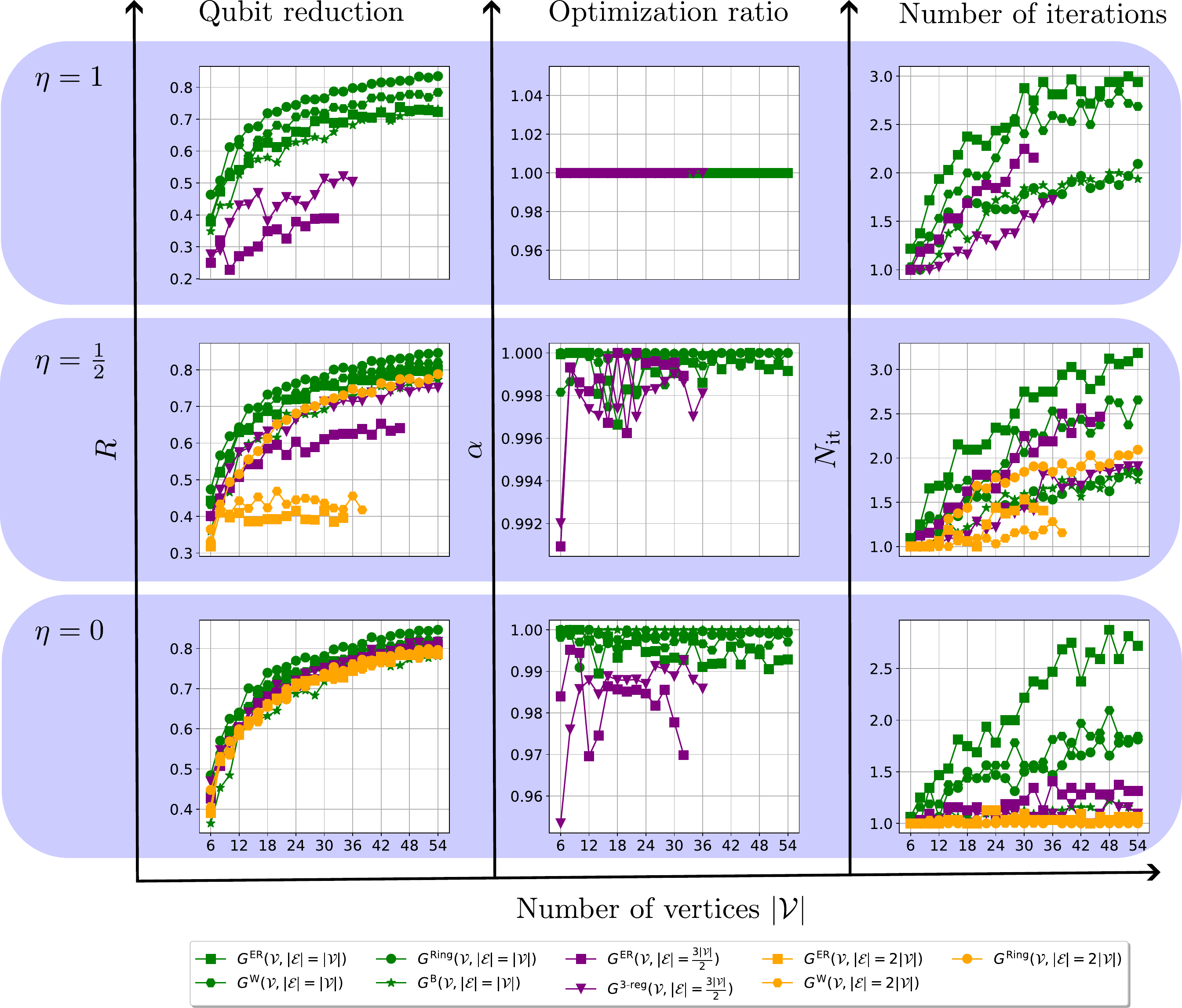}
\caption{Performance of the algorithm for $O_1=O_2=$brute force for a set of various random graphs as indicated in the legend (see Sec. \ref{sec:random_graphs} for further explanations). Left column: qubit reduction $R$ as defined in Eq. \eqref{eq:qubit_red}. Middle column: approximation ratio $\alpha$ as defined in Eq. \eqref{eq:approx_ratio}. Right column: number of iterations in the algorithm $N_{\text{it}}$. Each row shows results for retaining different fractions $\eta$ of the energy ranges in the subsystems, where $\eta=0$ means that only the degenerate ground states are retained, where another optimization $O_2$ is still performed. Green lines correspond to graphs with degree 2, purple lines to graphs with degree 3 and orange lines to graphs with degree 4 (The ground state energy was taken to be the energy found when $\eta=1$).}
\label{fig:brute_force}
\end{figure*}
In this section, we present numerical simulations of our \cref{alg:approach} for problem instances that have quadratic Hamiltonians as in Eq. \eqref{eq:two_body_ham2} and are representable on graphs. This comprises a wide category of problems with diverse structures. Thus, we consider various classes of weighted random graphs with different underlying structures, as described in \cref{sec:random_graphs}, where the underlying average degree of the vertices ranges from two to four. The metrics that we are primarily interested in are, on the one hand, the qubit reduction
\begin{equation} \label{eq:qubit_red}
    R := 1-\frac{N_q}{|\mathcal{V}|}
\end{equation}
where $N_q$ is the number of qubits required by our approach and $|\mathcal{V}|$ the number of variables in the original problem. This quantity takes values in $[0,1)$ and the closer the values are to 1, the better the reduction in the number of qubits. In addition, we explore the quality of the solutions that we found, as quantified by the approximation ratio,
\begin{equation} \label{eq:approx_ratio}
    \alpha := \frac{\tilde{E}_{0}}{E_{0}} ,
\end{equation}
where $\tilde{E}_{0}$ is the energy approximation of our approach and $E_0$ is the actual exact energy of the ground state. It is also interesting to consider the number of iterations $N_\mathrm{it}$ that the algorithm uses for a given problem instance. Furthermore, we are interested in how these metrics change when we reduce the size of the local subspaces considered by tuning the hyperparameter $\eta$, see Eq. \eqref{eq:eta-def}. In all simulations the clustering of graphs is done via the well known Louvain algorithm \cite{Louvain}, see \cref{sec:Louvain} for details. 

First, we run our algorithm in conjunction with a simulation of the Filtering Variational Quantum Eigensolver (F-VQE) \cite{filtering,filtering_case_study} as the optimizer of choice, i.e. $O_1=O_2=\text{F-VQE}$, while using the inverse filter. A short description can be found in the appendix of \cite{Schmid_2025}. This algorithm has the advantage that the circuit does not depend on the Hamiltonian, which is relevant when used as $O_2$ since other approaches, such as, e.g. QAOA, might require multi-qubit gates that need to be decomposed into single-and two-qubit gates when run on actual hardware. 
Yet, it is not our intention to argue that F-VQE would be the best suited optimizer for our algorithm. 
Moreover, since our simulations are a proof of principle, we do not include models of error channels and allocate a generous sampling budget of 1000 shots per circuit for instances with more than 14 vertices. In all simulations, we calculate the coupling array $\chi$ and each data point is an average of 32 random instances.

The results of these simulations for three different local energy ranges are shown in \cref{fig:proof_of_principle}. Even, when using the full energy range $\eta=1$ we notice a significant qubit reduction, e.g. a $\sim 45 \%$ reduction for weighted random 3-regular graphs on 24 vertices. The magnitude of reduction is clearly larger for graphs with lower average degree, which is explained by the increasing interaction strength $\normsmall{H_i^{(I)}}$ of the subgraphs, thus increasing the number of states that need to be considered for the global ground state. The exception is the ring graph connected to the nearest and next nearest neighbors. We attribute this behavior to the local structure of the graph, since the number of inter-community edges is six for a community of neighboring vertices, independent of the system size. 

Furthermore, an important observation is the increasing reduction with system size, which is a strong indicator of a growing benefit from our approach when considering larger systems. We also observe an increasing number of iterations used by the algorithm that hints at a connection between the hierarchical structure of the algorithm and the increasing reduction. The approximation ratio $\alpha$ is, as expected, almost $100 \%$, because the algorithm retains the global ground state. The small deviations are related to the performance of the optimizer.  

When reducing the energy range considered in the subsystems to half, $\eta= \frac{1}{2}$, we observe a stronger reduction for all graphs. As an example, we observe a $\sim67\%$ reduction for weighted random 3-regular graphs on 24 vertices compared to $\sim45\%$ for the same instances using $\eta=1$. In particular, graphs with an average degree of four get a sizable boost in reduction. These stronger reductions are especially relevant since the search space grows exponentially with the number of qubits $N_q$. However, they come with the caveat that there is no guarantee of retaining the global ground state. Nonetheless, we observe that the average optimization ratio for most graphs remains above $99.5\%$, indicating that very good solutions are found even at very strong qubit reduction. 

Reducing the energy range further, we only consider the degenerate local ground states, i.e. $\eta = 0$, while still optimizing in the reduced system via the optimizer $O_2$. This results in an amplification of the previously mentioned effects, increased reduction in qubit number and a reduced $\alpha$. Note that the qubit reduction for degree two graphs is not much better compared to the $\eta=\frac 1 2$ case because we are at the natural resolution of the Louvain algorithm. In other words, the qubits needed for the local systems become the bottleneck. 

In \cref{fig:brute_force} we consider larger system sizes. To avoid the need to classically simulate the quantum optimization algorithm, we simply used a brute force optimization for $O_1$ and $O_2$, which scans the eigenvalues for the lowest one. Here we obviously use the value of $E_0$ obtained by running our algorithm with $\eta=1$ as the exact ground state energy. The main observation is the continued increase in reduction for most graphs, e.g. a $\sim 75\%$ reduction for weighted random 3-regular graphs with $\eta=\frac{1}{2}$. This again indicates that larger reductions can be expected from our algorithm with growing system size.  

\begin{figure}[!t]
\centering
\includegraphics[scale=0.5]{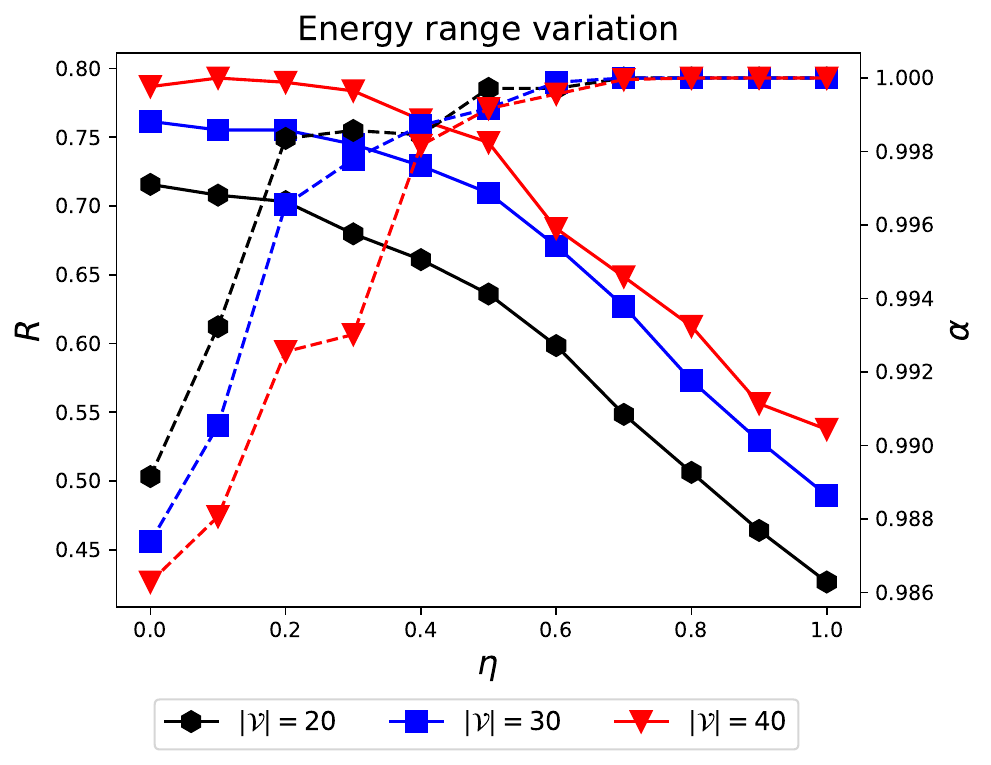}
\caption{Qubit reduction $R$ (solid lines) and approximation ratio $\alpha$ (dashed lines) for weighted random 3-regular graphs of different sizes ($|\mathcal{V}|=20,30$ or $40$) as a function of the fraction of retained energy range $\eta$. Here brute force optimization was used and the ground state was taken to be the energy found when $\eta=1$.}
\label{fig:eta_variation}
\end{figure}

In \cref{fig:eta_variation} we further analyze the relationship between $\eta$, $\alpha$ and $R$. To this end, we simulate weighted random 3-regular graphs for varying $\eta$ values using brute force optimization to ensure that no effects of  the optimizer affect the optimization value $\alpha$. For the system sizes considered, we see that the reduction increases significantly when reducing $\eta$. For the 40 vertices cases, the reduction goes from $\sim53\%$ for $\eta=1$ to $\sim75\%$ for $\eta=\frac{1}{2}$, while retaining an optimization ratio of $\sim99.9\%$. Decreasing $\eta$ further results in a greater reduction, but also impacts $\alpha$. For the system sizes considered, the value of $\alpha$ for $\eta=0$ is still relatively large, but one already sees a clear separation for different system sizes, indicating worsening optimization rations for larger systems. On the other hand $\eta=\frac{1}{2}$ appears to be a sweet spot that balances significant reduction with good approximation. For a short investigation regarding this behavior, see \cref{sec:comments_on_eta}. 

\section{Generalization to PUBOs} \label{sec:extensions}
A more general class of binary optimization problems is the class of polynomial unconstrained binary
optimization problems (PUBOs) with the Hamiltonian to optimize given by 
\begin{equation}\label{eq:Pubo_hamiltonian}
    H=\sum_{S\subseteq [n]}J(S)\prod_{j\in S}Z_j.
\end{equation}
Although a PUBO can be reduced to a QUBO \cite{Pubo_to_Qubo}, it requires a significant amount of auxiliary qubits and some optimization tasks are better suited to the higher order version \cite{Pubo_better_than_Qubo}. Our reduction approach can with some slight generalizations be natively used for PUBOs.  
We represent the Hamiltonian of Eq. \eqref{eq:Pubo_hamiltonian} as a hypergraph with $n$ vertices, c.f. \cref{subsec:pubo_hypergraphs}. Hypergraphs can be clustered similar to graphs using a suitable modularity extension, see \cite{h-clusterin-comp} for a comparison of modularity-based approaches and \cite{h-lovain} for an adaptation of the Louvain algorithm called h-Louvain. After the clustering, the Hamiltonian is again written in terms of community Hamiltonians and inter-community hyperedges, 
\begin{align}
    H&=\sum_{i=1}^N H_i^{(C)}+\sum_{S\subseteq [n]: S\nsubseteq \Com_i\,\forall i}J(S)\prod_{j\in S}Z_j\\
    H_i^{(C)}&=\sum_{S\subseteq \Com_i}J(S)\prod_{j\in S}Z_j
\end{align}
Using the same argument as in \cref{subsec:Hilber_space_reduction}, we can group every inter-community hyperedge containing at least one vertex in $\Com_i$ into an interaction Hamiltonian $H_i^{(I)}$. Formally, the set of these hyperedges can be defined as
\begin{equation}
    \mathcal{E}_i^{(I)}:=\{S\subseteq[n]: S\nsubseteq \Com_i \land \exists v_j: v_j\in \Com_i\cap S \} .
\end{equation}
All hyperedges that contain no vertex of $\Com_i$ are then part of $H_\mathscr{E}$, allowing for a generalized version of \cref{thm:cut_off_two_body}.
\begin{theorem}\label{thm:cut_off_general}
Given a Hamiltonian $H$ as in Eq. \eqref{eq:Pubo_hamiltonian} and its decomposition into $\Com_i$ and the environment as in \eqref{eq:decomp_environment}, a local state $\ket{\psi_i^\mu}$ with $H_i^{(C)}\ket{\psi_{i}^{\mu}} = E^{\mu}_i \ket{\psi_{i}^{\mu}}$, can only be part of the global ground state if its local energy satisfies $E_i^\mu\in [E_i^0, E_i^0+\Delta_i]$, where the spectral range $\Delta_i$ is given by
\begin{equation}\label{eq:Pubo_cut-off}
    \Delta_i=\lambda_{i,\mathrm{max}}^{(I)}-\lambda_{i,\mathrm{min}}^{(I)}\leq 2\norm{H_i^{(I)}},
\end{equation}
with $\lambda_{i,\mathrm{min}}^{(I)}$ and $\lambda_{i,\mathrm{max}}^{(I)}$ being the smallest and largest eigenvalues of the interaction Hamiltonian.\\
\end{theorem}
The proof of this theorem is given in \cref{sec:cut_off_proof}. The operator norm of the interaction Hamiltonian can be bounded by the sum of absolute values of inter-community hyperedge weights.
\begin{equation}
    \norm{H_i^{(I)}} \leq \sum_{e\in \mathcal{E}_i^{(I)}} |J(e)|.
\end{equation}
After solving all subsystems with a suitable PUBO optimizer, building an optimization problem in a reduced Hilbert space is done analogously to \cref{subsec:reduced_hamiltonian}. The reduced local Hamiltonian $\tilde{H}_i^\mathrm{(C)}$ is again constructed by an energy ordering as in Eq. \eqref{eq:reduced_community_hamiltonian}. The Hamiltonians of the interaction edges $\tilde{H}_e$ for $e\in \mathcal{E}^{(I)}$ ($\mathcal{E}^{(I)}=\bigcup_i \mathcal{E}_i^{(I)}$) are first handled individually. To this end, let $\vec{c}(e)$ denote all community indices whose vertices are part of $e$. Then the diagonal elements are products of the weights and the corresponding $\chi$ values,
\begin{equation}
    \tilde{H}_e \bigotimes_{i\in\vec{c}(e)}\ket{\psi^{\mu}_i}=J(e)\chi_{e}{(\vec{\mu})}\bigotimes_{i\in\vec{c}(e)}\ket{\psi^{\mu}_i} \, .  
\end{equation}
With shorthand notation for the indices that describe the labeling of local states $\vec{\mu}=\{\mu(i)\}_{i\in \vec{c}(e)}$ and $\ket{\Psi}:=\bigotimes_{i\in\vec{c}(e)}\ket{\psi^{\mu}_i}$, the coupling sign reads
\begin{equation}
\chi_{e}(\vec{\mu})=\bra{\Psi} \prod_{j\in e} Z_j\ket{\Psi} 
\end{equation}
If multiple hyperedges, e.g. $e_i$ and $e_j$, span the exact same communities $\vec{c}(e_i)=\vec{c}(e_j)$, they can be reduced again to a single hyperedge by summation of the diagonal elements. Note that the number of $\chi$ values for a single edge $N_\chi(e)$, which needs to be calculated individually, if the Hamiltonian is required for gate sequences, is given by
\begin{equation}
    N_\chi(e)=\prod_{i \in \vec{c}(e)}\tilde{d}_i.
\end{equation}
Going to higher iterations is completely analogous to the procedure described in \cref{subsec:iteration} with the adapted $\Delta_i$ of Eq. \eqref{eq:Pubo_cut-off} in mind.

\section{Conclusion and Outlook}
We have introduced a hierarchical algorithm for combinatorial optimization problems that 
significantly reduces the number of qubits and the configuration space size required to solve the problem. This is achieved by representing the problem as a graph and splitting the system into subgraphs with strong internal interaction and weak links to other subgraphs. For such splittings, we determine an effective energy range for low lying local states that guarantees the conservation of the ground state. Once this relevant subspace is found, we recombine the problem in a reduced form using a binary encoding with a local energy ordering. This procedure can be iterated until the remaining problem can be solved in a single step. 

In numerical simulations, we tested this approach for a diverse set of sparse weighted random graphs up to an underlying average vertex degree of 4 and a simulation of the filtering variational quantum eigensolver \cite{filtering} to give a working proof of principle example. Moreover, we found that the qubit reduction increases significantly when reducing the retained energy range in each subgraph by a factor $\eta < 1$ while the approximation ratio is only slightly impacted for a large tuning range. For the example of weighted random 3-regular graphs with 40 vertices, the reduction goes from $\sim53\%$ for $\eta=1$ to $\sim75\%$ for $\eta=\frac{1}{2}$, while retaining an approximation ratio of $\sim99.9\%$. We also find an increased reduction with growing system size, which is essential for scaling. Furthermore, we generalized the ansatz to multi body Hamiltonians, or equivalently PUBO problems, where the strict local energy range to guarantee ground state conservation was found to be larger. 

In future work, it would be interesting to see which optimizers are best suited for the different optimization tasks inside the algorithm.  
It would also be interesting to see whether the strict local energy range for PUBO problems can be reduced by a larger factor $\eta^\mathrm{PUBO}$ than the quadratic version, so that the effective ranges would be similar. Furthermore, the impact on the approximation ratio when excited states are included for large and dense graphs as studied in \cite{zhou2023qaoa} is worth investigating. Lastly, it would be desirable to specify a criterion or measure that determines how much a problem is suited for our ansatz.

\begin{acknowledgments}
The authors thank Jens Eisert, Maximilian Kramer and Franz Schreiber for fruitful discussions.
This work is part of the Munich Quantum Valley, which is supported by the Bavarian state government with funds from the Hightech Agenda Bayern Plus. It also received funding from the German Federal Ministry of Education and Research via the funding program "quantum technologies - from basic research to the market" under contract number 13N16182 "MUNIQC-SC".
\end{acknowledgments}

\section*{Data Availability}
Data and employed simulation codes are available upon request.

\newpage
\appendix

\section{Combinatorial Optimization on Quantum Computers}\label{sec:app-graphs}

This chapter explains the relation between combinatorial optimization tasks, the reformulation as ground state problems of Hamiltonians, and their representation as graphs.

A combinatorial optimization problem is defined by a pair \( (\mathcal{X},C)\), where $\mathcal{X}$ denotes a countable set of solutions and $C:\mathcal{X}\rightarrow\mathds{R}$ is a cost function. The goal of the optimization process is to find a $x\in \mathcal{X}$ s.t. $C(x)\leq C(y)$, $\forall y\in \mathcal{X}$. In particular, most often $\mathcal{X}$ is finite and solutions can w.o.g. be represented by bit strings of length $n$, i.e. $\mathcal{X}=\{0,1\}^n$ and the set of possible cost functions is the set of pseudo boolean functions $\mathcal{R}_n=\{C:\{0,1\}^n\rightarrow\mathds{R}\}$.
Since the Boolean functions on $n$ bits $\mathcal{B}_n=\{C:\{0,1\}^n\rightarrow\{0,1\}\}$ form a basis for $\mathcal{R}_n$,
one often writes,
\begin{equation}
    C(x)=\sum_{\alpha=1}^m J_\alpha C_\alpha (x),
\end{equation}
with $C_\alpha:\{0,1\}^n\rightarrow\{0,1\}$. In many satisfaction problems, where the $C_\alpha$ are boolean clauses and $J_\alpha\in\{0,1\}$, those clauses depend on only a few of the $n$ bits \cite{QAOA} . Since we want to use quantum computers to solve such problems, it is useful to convert these classical cost functions into Hamiltonians.

The space of cost functions $\mathcal{R}_n$ is isomorphic to the space of $2^n \times 2^n$ diagonal real matrices, i.e. diagonal Hamiltonians acting on $n$ qubits and we examine the transformation to such representing Hamiltonians in the following, compare \cite{Boolean_Hamilton}.
A Hamiltonian represents a function $f$ if it satisfies all $2^n$ eigenvalue equations.
\begin{equation}
    \forall x \in \{0,1\}^n\quad H_f\ket{x}=f(x)\ket{x}
\end{equation}
An example of a representation of a $n$-bit real valued function is given by
\begin{equation}
    \prod_{j\in S}Z_j\ket{x}=\chi_S(x)\ket{x},
\end{equation}
where $\chi_S(x):\{0,1\}^n\rightarrow\{-1,+1\}$ is the parity function that yields $+1$ iff the number of active bits of $x$, that are one, is even. $Z_j:=I\otimes I\otimes \cdots \otimes Z \otimes \cdots \otimes I$ denotes the Pauli $Z$ acting on the $j$-th qubit, and $S$ is a subset of active qubits $S\subset [n]:=\{1,2,\cdots,n\}$.
The set of parity functions $\{\chi_S(x):S\subset[n]\}$ is also a basis of the $n$-bit real functions $\mathcal{R}_n$. Thus, we can write for a Boolean function $f\in\mathcal{B}_n$ 
\begin{equation}
    f(x)=\sum_{S\subset [n]}\hat{f}(S)\chi_S(x).
\end{equation}
With this a representing Hamiltonian for any Boolean function can be written as,
\begin{align}\label{eq:ham_conversion}
    H_f&=\sum_{S\subset [n]}\hat{f}(S)\prod_{j\in S}Z_j\\
    &=\hat{f}(\emptyset)I+\sum_{j=1}^n \hat{f}(\{j\})Z_j+\sum_{j<k}\hat{f}(\{j,k\})Z_jZ_k+\cdots \nonumber
\end{align}
where the expansion coefficients are given by,
\begin{equation}
    \hat{f}(S)=\frac{1}{2^n}\sum_{x\in\{0,1\}^n}f(x)\chi_S(x)=:<f,\chi_S>.
\end{equation}
Calculating such a representation is in general not tractable, however in the case of constraint satisfaction problems with clauses acting only locally on $k=\mathcal{O}(\log n)$ qubits, which are often considered as quantum computing use cases, an efficient implementation is possible.
For proofs and approaches to convert specific types of Boolean cost functions and their compositions to a representing Hamiltonian, see \cite{Boolean_Hamilton}. For many relevant problems, it is straightforward to write down the representing Ising Hamiltonians with at most quadratic terms \cite{lucas2014ising}.
\subsection{Two body Hamiltonians, QUBO and Ising model}\label{subsec:qubo_and_ising}
The main show-case for the algorithm presented in this paper is for problems in quadratic form, which only have two body interaction terms, i.e. functions that can be represented by Hamiltonians of the form,
\begin{equation}\label{eq:two_body_ham0}
    H_f=\sum_{a<b}J_{a,b}\, Z_a\, Z_b.
\end{equation}
These Hamiltonians are equivalent to the famous weighted maximum cut problem \cite{max_cut_quadratic}.
It is possible to represent the Hamiltonians in Eq. \eqref{eq:two_body_ham0} as graphs and consequently use graph theory to our advantage.

A graph $G=(\mathcal{V},\mathcal{E})$ is a pair of the set $\mathcal{V}$, which denotes all vertices of the graph, and the set $\mathcal{E}$, which denotes all edges of the graph. The elements of $\mathcal{E}$ are a pair of vertices $\{v_i,v_j\}$ connected through an edge. In addition, a set $W$ can be introduced that contains weights attached to the edges of the graph, making it a weighted graph. The degree of a vertex $\deg(v_i)$ is given by the number of edges incident to the vertex.

Given an optimization problem with $n$ input bits and a cost function represented by a Hamiltonian as in Eq. \eqref{eq:two_body_ham0}, we can associate the $n$ bits with the vertices of a graph. A term $J_{i,j}Z_iZ_j$ is converted into an edge $\{v_i,v_j\}$, which has a weight $w_{i,j}=J_{i,j}$ with $ w_{i,j}\in W$. Additionally, we can give the vertices $v_i$ a state attribute $s_i \in \{0,1\}$ that is useful to visually represent computational basis states and retrieve energy values from the graph representation.

The Ising model Hamiltonians \cite{Gallavotti99, lucas2014ising} can be written as,
\begin{equation}\label{eq:two_body_ham}
    H_f=\sum_{i<j}J_{i,j}Z_iZ_j+\sum_{i}h_i Z_i+cI.
\end{equation}
This is a widely studied model that is used to represent many famous NP-complete problems \cite{lucas2014ising}.
It is also equivalent to quadratic unconstrained binary optimization (QUBO) problems,
\begin{equation*}\label{eq:qubo}
\hspace{-0.4cm}
 \quad \min_{x\in\{0,1\}^n} x^T Q x+ c = \min_{x\in{0,1}^n}\sum_{i<j}Q_{i,j}x_ix_j+\sum_{i}Q_{i,i} x_i+ c \, ,
\end{equation*}
where $Q\in \mathds{R}^{n\times n}$ is an upper triangular matrix and the transformation is given by $Z_i=2x_i-1$.

In \cref{sec:cut_off_proof} we will proof different bounds on local energy cut-off ranges $\Delta_i$ for different types of problems. The Ising model Hamiltonians of Eq. \eqref{eq:two_body_ham} can be treated in three different ways together with our algorithm. Firstly, one can ignore the local fields in the clustering, see \cref{sec:clustering}, but then the larger bound in \cref{sec:cut_off_proof} applies. Secondly, it is possible to map Ising Hamiltonians with diagonal terms, therefore also QUBO problems, to purely quadratic Hamiltonians by introducing an auxiliary variable that is connected to all vertices with a local field \cite{Barahona1989}, then the two-body cut-off is applicable. Lastly, one can choose to treat the Ising model with linear terms as a general polynomial unconstrained binary optimization (PUBO) problem.
\subsection{PUBO and hypergraphs}\label{subsec:pubo_hypergraphs}
The polynomial unconstrained binary optimization (PUBO) problem is the optimization of a pseudo-Boolean function $f:\{0,1\}^n \rightarrow \mathds{R}$, which has the general form of a pseudo-Boolean polynomial \cite{pseudo_boolean_functions}. The representative Hamiltonian can be written as
\begin{align}\label{eq:PUBO_ham_appendix}
    H_f&=cI+\sum_i^nJ_iZ_i+\sum_{i<j}^n J_{i,j}Z_iZ_j+\dots\nonumber\\
    &=\sum_{S\subseteq [n]}J(S)\prod_{j\in S}Z_j
\end{align}
A hypergraph $H(\mathcal{V},\mathcal{E})$ is a generalization of the graph $G(\mathcal{V},\mathcal{E})$, in which the edges $e_i \in \mathcal{E}=\{e_i\}_{i\in I}$ are subsets of $\mathcal{V}$ with arbitrary cardinality, called hyperedges. Here $I$ is a finite index set and every edge $e_i$ can be given a weight $w_i$.
\begin{figure}[!ht]
\centering
\includegraphics[scale=0.07]{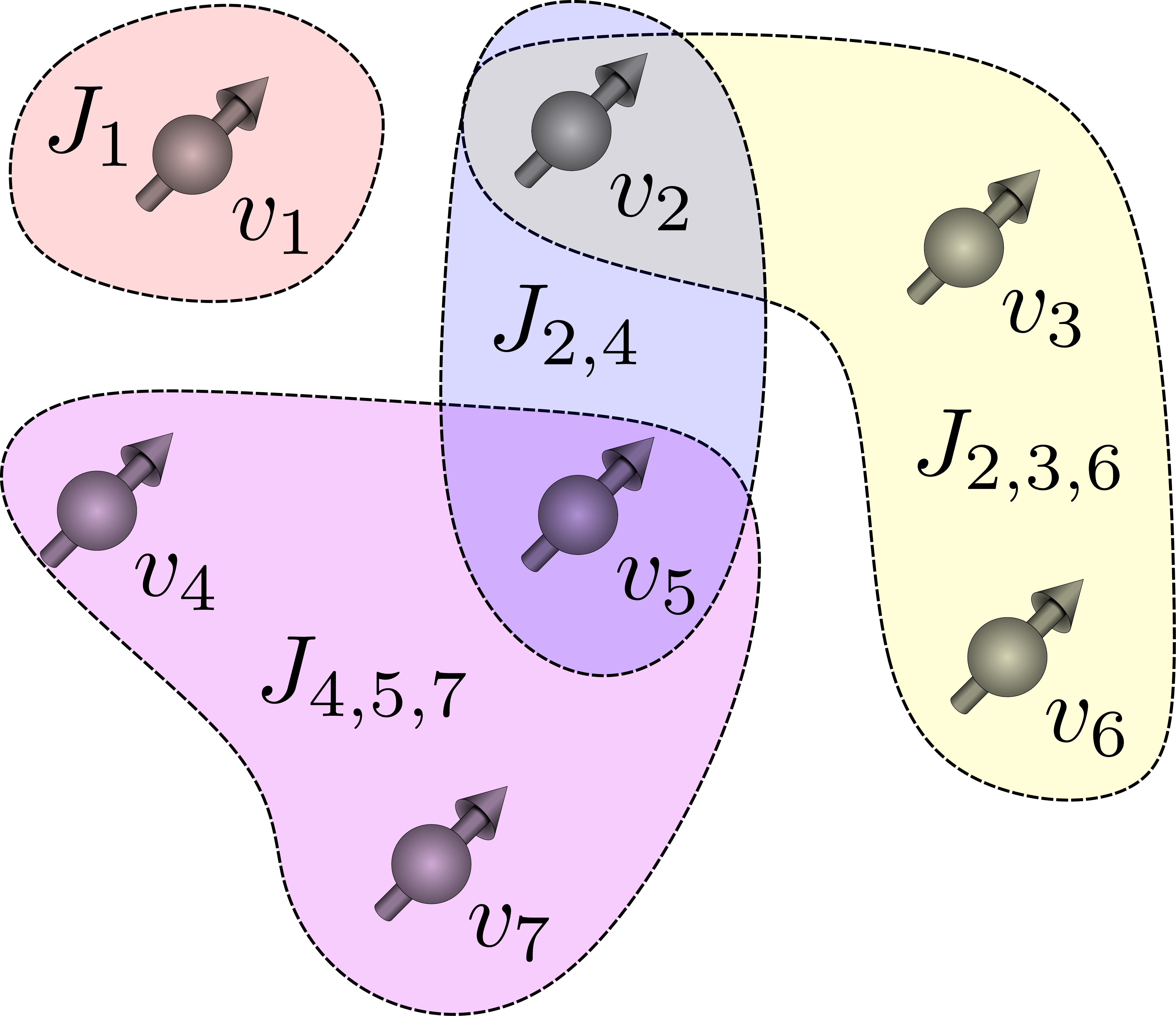}
\caption{A visualization of the hypergraph associated to an example Hamiltonian $H=J_1Z_1+J_{2,4}Z_2Z_5+J_{2,3,6}Z_2Z_3Z_6+J_{4,5,7}Z_4Z_5Z_7$. Hyperedges are depicted by the shaded regions and represent multi-body terms.}
\label{fig:hypergraph}
\end{figure}

The transformation from Eq. \eqref{eq:PUBO_ham_appendix} to a hypergraph is given by the vertex set $\mathcal{V}=\{v_1,\dots,v_n\}$ and the edges $e_i=\{v_i\}_{i\in S}$ with $w_i=J(S)$ for all $S$ s.t. $J(S)\neq0$. Again, an attribute $s_i\in\{0,1\}$ can be attached to the vertices to represent a computational basis state. An example is visualized in \cref{fig:hypergraph}.

\section{Clustering}\label{sec:clustering}
After mapping our problem to a graph, we want to divide the resulting graph into subsystems. In this context, we will discuss communities of graphs and a method to detect them. For an extension to hypergraphs, see \cite{h-clusterin-comp} and \cite{h-lovain}.
\subsection{Communities}\label{sec:community}

\begin{figure}[!ht]
\centering
\includegraphics[width=\columnwidth]{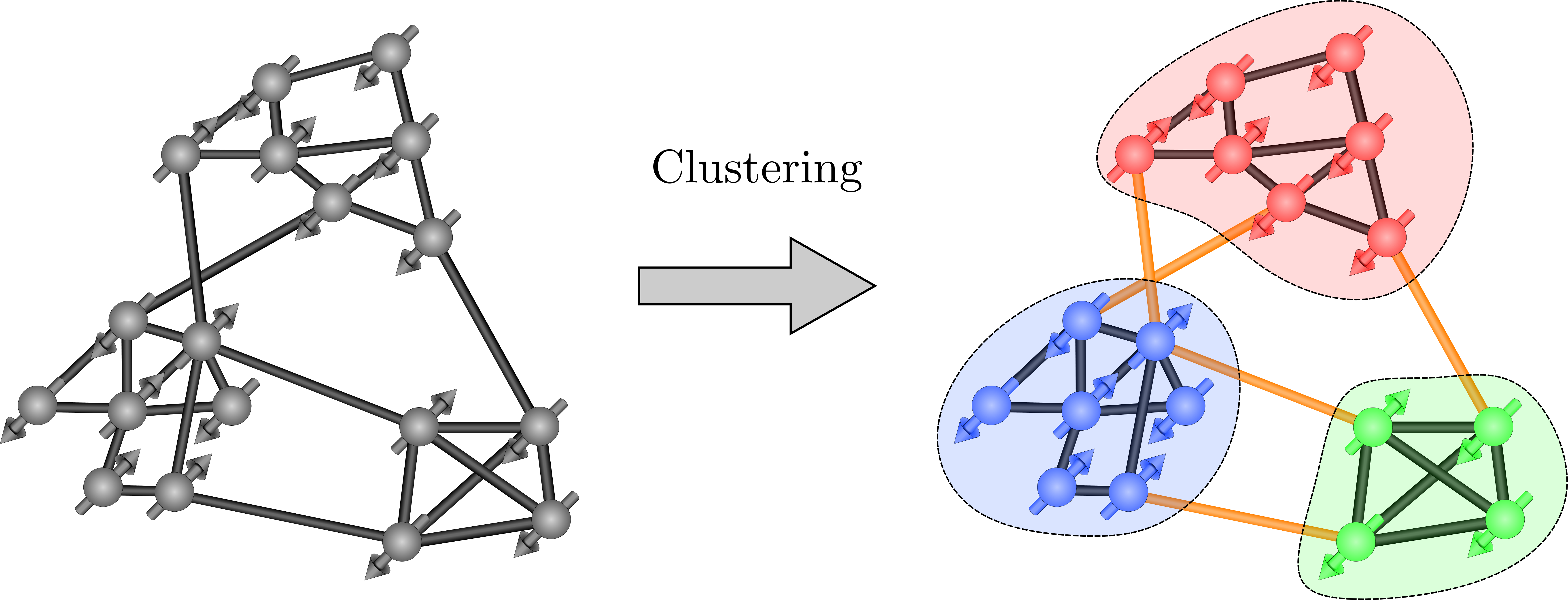}
\caption{The graph shown reveals a community structure, the three shaded groups of nodes have dense connections within the group while the groups are only sparsely connected to each other.}
\label{fig:community}
\end{figure}

Many graphs tend to have sets of vertices which group together and act relatively independently from other vertices in the graph.  
These tendencies are captured by dividing the graph into disjoint subsets, called communities. The vertices within a subset are densely connected, while the communities have only a few connections to each other. The union of all communities is the whole graph. An example of a graph showing the community structure is given in \cref{fig:community}. 
This general understanding of what a community is supposed to be can lead to many different definitions of the term. However, many algorithms entirely forgo a concrete definition. The community detection algorithm discussed in \ref{sec:Louvain} is based on modularity optimization. 
\begin{figure*}[!t]
\Centering
\includegraphics[scale=0.0375]{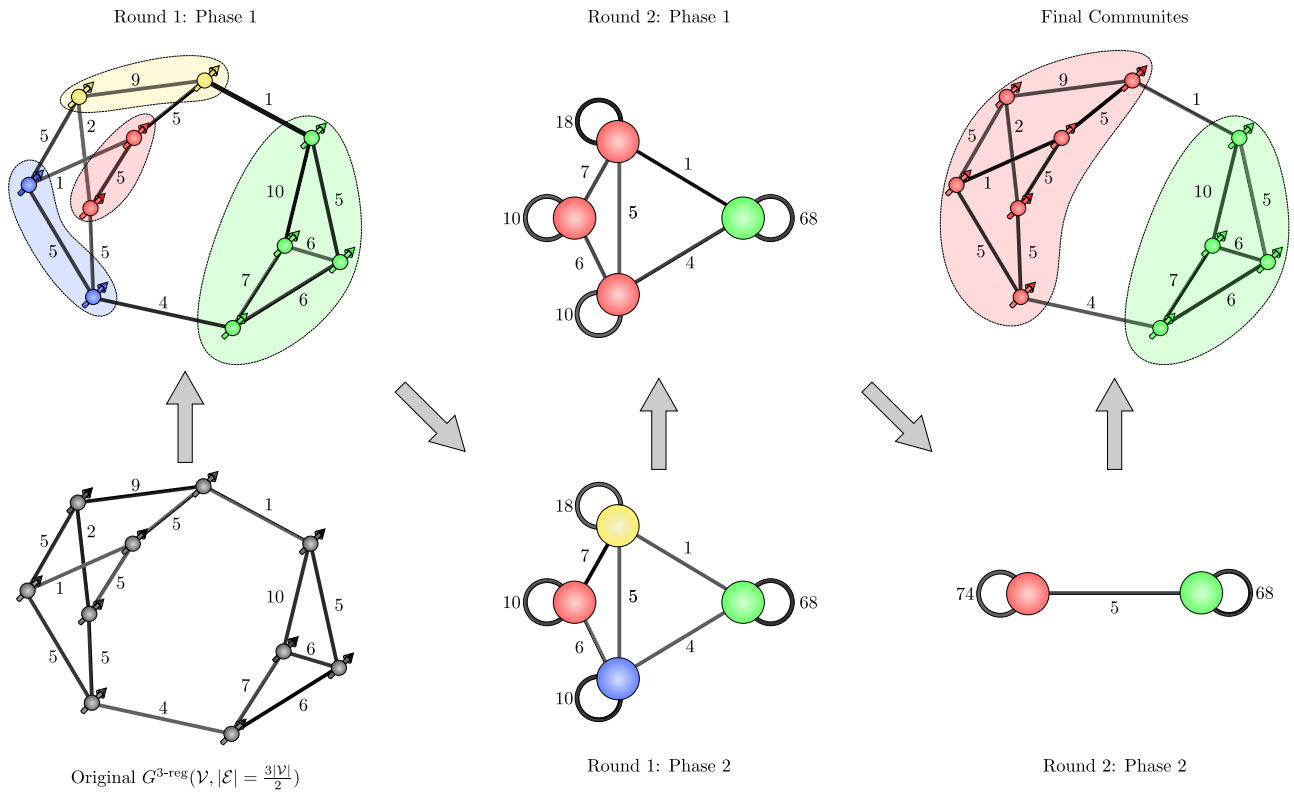}
\caption{Visual representation of the Louvain algorithm. In the first phase modularity is optimized by the local action of moving nodes to neighboring communities. In the second phase communities are contracted to single nodes, while edges between communities are summed up. Intra-community edges become self loops and their weight is doubled to preserve the correct modularity. The process is repeated until no gain in modularity can be made.}
\label{fig:Louvain}
\end{figure*}
\subsection{Modularity}
The measure of modularity was first introduced in  \cite{Modularity_first} and is based on a previous measure in \cite{Modualrity_before}, see also \cite{Modularity_used_to_explain}. It assigns a score to a division of the graph and is defined as the number of edges falling within a community minus the expected number, based on a network with the same edge and node count where the edges are distributed at random, multiplied by a normalization factor. If weighted edges are included, modularity is given by \cite{Louvain},
\begin{equation}
    Q=\frac{1}{2m}\sum_{i,j}\left[w_{i,j}-\frac{k_i k_j}{2m}\right]\delta(c(i),c(j)) .
\end{equation}
Here $w_{i,j}$ is the weight associated with the edge connecting the vertices $i$ and $j$, $k_i=\sum_j w_{i,j}$ is the sum of weights from all attached edges to vertex $i$, $c(i)$ is the community to which vertex $i$ is assigned, $\delta(c(i),c(j))=1 \text{ iff }c(i)=c(j)$ and $m=\frac 1 2 \sum_{i,j} w_{i,j}$ is the sum of all weights. Therefore, a value of $Q=0$ would not be better than to assign communities at random and values close to $Q=1$ indicate a strong community structure.
\subsection{Louvain algorithm}\label{sec:Louvain}
The Louvain method \cite{Louvain} is a heuristic method based on modularity optimization used to extract community structures of large graphs in a short computation time, and running the algorithm on a graph of 118 million nodes took only 152 min. In contrast, exact modularity optimization is computationally difficult \cite{exact_modularity_hard}. The algorithm has two phases, which are repeated until no further improvement can be made. 

For a graph $G=(\mathcal{V},\mathcal{E})$ with $N=|\mathcal{V}|$ nodes, one starts by assigning one community to each node. Now, for a node $v_i$ the change in modularity is calculated when it is removed from its community and instead placed in a neighboring community. If there is a positive change, this node is placed in the community that yields a maximum gain in modularity. This is repeated for all nodes and then again repeated in sequence until no further improvement can be made by an individual move. This finishes phase one, where the efficiency of the algorithm is partly due to the fact that the modularity change induced by moving one node to a neighboring community can be easily computed. 

In the second phase, the communities found are contracted and become the nodes of a new graph. The weight of the edges between two new nodes is the sum of the weights associated with the edges connecting the original communities. The inner edges of the community become self-loops of the new nodes, and the weight is doubled to account for summing over it just once. These two phases are repeated until there are no changes in modularity. A visual representation is shown in \cref{fig:Louvain}.

Since the network becomes increasingly smaller in each iteration, most of the computation time is spent in the first iteration. In addition to the advantage of fast computation time,
this method produces a hierarchical structure of the network, and it is possible to extract communities of low scale by looking at the earlier iterations, which is typically hard for modularity-based approaches \cite{Resolution_limit_modularity}.

However, it should be noted that the use of this clustering method combined with our approach in \cref{sec:approach} produces the qubit reductions shown in \cref{sec:simulations} even for graphs without an inherent community structure, such as for the Erdős–Rényi model and the Barabasi model, see \cref{sec:random_graphs}, if they are of moderate degree. Furthermore, the algorithm only works for non negative edge weights, we thus use a graph with $w_{i,j}^\prime=|w_{i,j}|$ as weights during the clustering process, which produces meaningful partitions based on interaction strength.

\section{Proof of Energy Cut-off}\label{sec:cut_off_proof}
In this section, we present the proof of \cref{thm:cut_off_two_body} and \cref{thm:cut_off_general} of the main text. We start with a general Hamiltonian and later consider Hamiltonians that purely consist of two-body terms, for which we find a tighter bound.
\subsection{General cut-off: proof of \texorpdfstring{\cref{thm:cut_off_general}}{theorem 2}}
\begin{proof}
Given an arbitrary cost function Hamiltonian $H$ acting on $n$ qubits that are represented by $\mathcal{V}$, compare Eq. \eqref{eq:Pubo_hamiltonian_main}, we consider its decomposition into community a $\Com_i\subset \mathcal{V}$ and treat the remaining system $\mathscr{E}:=\mathcal{V} \setminus \Com_i$ as an environment. The terms of $H$ that consist only of Pauli-Zs acting on qubits in $\Com_i$ are gathered in $H_i^{(C)}$, while the terms only acting on $\mathscr{E}$ are combined into $H_\mathscr{E}$, and the remaining terms that act on both are written as an interaction Hamiltonian $H_i^{(I)}$,
\begin{equation}\label{eq:splitting_com_env}
    H = H_i^{(C)}+ H_i^{(I)} + H_\mathscr{E} .
\end{equation}
We consider the product state $\ket{\psi^0_i,\psi_\mathscr{E}^0}$ of the local community ground state ($H_i^{(C)}\ket{\psi^0_i} = E^0_{i}\ket{\psi^0_i}$) and the ground state of the environment ($H_\mathscr{E} \ket{\psi^0_\mathscr{E}} = E^0_{\mathscr{E}}\ket{\psi^0_\mathscr{E}}$), and bound its energy by
\begin{equation}\bra{\psi^0_i,\psi_\mathscr{E}^0}H\ket{\psi^0_i,\psi_\mathscr{E}^0}\leq E^0_{i}+\lambda_{i,\mathrm{max}}^{(I)}+ E^0_\mathscr{E},
 \end{equation}
where $\lambda_{i,\mathrm{max}}^{(I)}$ is the largest eigenvalue of $H_i^{(I)}$. Note here that $\ket{\psi^0_i,\psi_\mathscr{E}^0}$ is an eigenstate of $H$, just as all computational basis states. As a consequence, the global ground state energy is also bounded by
\begin{equation}\label{eq:ground_state_bound_general}
    E^0\leq E^0_{i}+\lambda_{i,\mathrm{max}}^{(I)}+ E^0_\mathscr{E}.
\end{equation}
We now assume that a local state $\ket{\psi_{i}^{\mu}}$ of $\Com_i$ with $H_i^{(C)}\ket{\psi_{i}^{\mu}}=E_{i}^{\mu}\ket{\psi_{i}^{\mu}}$ is part of the global ground state, then there must exist a state of the environment $\ket{\psi_\mathscr{E}^{\nu}}$ with $H_\mathscr{E}\ket{\psi_\mathscr{E}^{\nu}}= E_\mathscr{E}^\nu \ket{\psi_\mathscr{E}^{\nu}}$ s.t. their total energy is not larger than this bound.
\begin{equation}E^\mu_{i}+\bra{\psi_i^\mu,\psi_\mathscr{E}^\nu}H_i^{(I)}\ket{\psi_i^\mu,\psi_\mathscr{E}^\nu}+E_\mathscr{E}^\nu\leq E^0_{i}+\lambda_{i,\mathrm{max}}^{(I)}+E^0_\mathscr{E}
    \end{equation}
We now use that $E^0_\mathscr{E}-E_\mathscr{E}^\nu \leq 0$ and
$
-\bra{\psi_i^\mu,\psi_\mathscr{E}^\nu}H_i^{(I)}\ket{\psi_i^\mu,\psi_\mathscr{E}^\nu}\leq - \lambda_{i,\mathrm{min}}^{(I)}$, 
where $\lambda_{i,\mathrm{min}}^{(I)}$ is the smallest eigenvalue $H_i^{(I)}$, to get a condition for the local energy:
\begin{equation}
    E^\mu_{i}\leq E^0_{i}+\lambda_{i,\mathrm{max}}^{(I)}-\lambda_{i,\mathrm{min}}^{(I)}=:E^0_i+\Delta_i
\end{equation}
Therefore, local eigenstates $\ket{\psi_{i}^{\mu}}$ with $E_i^{\mu} > E_{i}^0+\Delta_i$ cannot be part of the global ground state $\ket{\psi_0}$, because there is not enough energy in the interactions to shift them to low enough energies.
\end{proof}
\subsection{Two-Body cut-off: proof of \texorpdfstring{\cref{thm:cut_off_two_body}}{theorem 1}}
We now show that for Hamiltonians that exclusively consist of two-body terms, a tighter bound exists on the energy range $\Delta_i$, for which eigenstates of $H_i^{(C)}$ can contribute to the global ground state.
\begin{proof}
We consider Hamiltonians that exclusively consist of two-body terms.
\begin{equation}\label{eq:even_hamiltonian}
    H=\sum_{a<b}J_{a,b}\, Z_a\, Z_b
\end{equation} 
Such Hamiltonians have a $\mathds{Z}_2$ symmetry, as can be seen
when considering each summand $Z_a\, Z_b$ individually. Since 
\begin{equation}
    X^{\otimes N} Z_a\, Z_b X^{\otimes N} = Z_a\, Z_b
\end{equation}
flipping all bits of a state leaves its energy invariant,
\begin{equation}
     \bra{\psi^\mu}X^{\otimes N} H X^{\otimes N}\ket{\psi^\mu}=\bra{\psi^\mu} H \ket{\psi^\mu} .
\end{equation}
This property also holds for individual subsystems and induces a degeneracy in their ground states since flipping all local qubits leaves their Hamiltonian invariant,
\begin{equation}\label{eq:local_symmetry}
    X^{\otimes N_i} H^{(C)}_{i} X^{\otimes N_i} = H^{(C)}_{i} .
\end{equation}
On the other hand, the interaction Hamiltonian $H^{(I)}_{i}$ also consists of terms $Z_a\, Z_b$, where, however, one vertex is always within the community while the other is in the environment. In this case, flipping all qubits within the community yields a minus sign,
\begin{equation}\label{eq:local_anti_symmetry}
    X^{\otimes N_i} H^{(I)}_{i} X^{\otimes N_i} = - H^{(I)}_{i} .
\end{equation}
We now consider the product states $\ket{\psi_i^0,\psi_\mathscr{E}^0}$, $X^{\otimes N_i}\ket{\psi_i^0,\psi_\mathscr{E}^0}$ and use the properties of Eq. \eqref{eq:local_symmetry} and Eq. \eqref{eq:local_anti_symmetry} to obtain an expression for their energies
\begin{equation}
    \bra{\psi^0_i,\psi_\mathscr{E}^0}H\ket{\psi^0_i,\psi_\mathscr{E}^0}=E^0_{i}+\lambda_{i}^{(I)}+ E^0_\mathscr{E}
    \end{equation}
  and
  \begin{equation}
    \bra{\psi^0_i,\psi_\mathscr{E}^0}X^{\otimes N_i} H X^{\otimes N_i}\ket{\psi^0_i,\psi_\mathscr{E}^0}= E^0_{i}-\lambda_{i}^{(I)}+ E^0_\mathscr{E}
\end{equation}
with $\lambda_{i}^{(I)}:= \bra{\psi_i^0,\psi^0_\mathscr{E}} H^{(I)}_{i} \ket{\psi_i^0,\psi^0_\mathscr{E}}$.
Since either $\lambda_{i}^{(I)}$ or $-\lambda_{i}^{(I)}$ is  smaller than zero, the ground state energy can be bounded by
\begin{equation}
    E^0\leq E^0_{i}+ E^0_\mathscr{E}.
\end{equation}
This forms a reduced bound as compared to Eq. \eqref{eq:ground_state_bound_general}, from which a two-body cut-off arises following completely analogous arguments,
\begin{equation}\label{eq:two_body_cutoff_proof}
    E^\mu_{i}\leq E^0_{i}-\lambda_{i,\mathrm{min}}^{(I)}=:E^0_i+\Delta_i .
\end{equation}
Also note that Eq. \eqref{eq:local_anti_symmetry} implies that $\lambda_\mathrm{max}^{(I)}=-\lambda_\mathrm{min}^{(I)}=\normsmall{H^{(I)}_{i}}$.
\end{proof}
When iterating the algorithm, the reduced system Hamiltonian is again clustered so that it can be written in the form of Eq. \eqref{eq:splitting_com_env}. Since all states which are acted upon by $H_l^{(C,\mathrm{it})}$ of Eq. \eqref{eq:iter_local_hamiltonian} are originally the low lying states of a quadratic Hamiltonian, it is clear that each state in the reduced basis is part of a pair of states that correspond to flipping all original local bits of the new community $\Com_l$, thus they have the same local energy. Since the iterated interaction Hamiltonian also carries the $ZZ$-interactions over from the original graph, these bit flipped states then again have mirrored interaction strength as in Eq. \eqref{eq:local_anti_symmetry}. The same arguments leading to Eq. \eqref{eq:two_body_cutoff_proof} follow.

\section{Random Graphs}\label{sec:random_graphs}
\begin{figure}[!ht]
\centering
\includegraphics[scale=0.04]{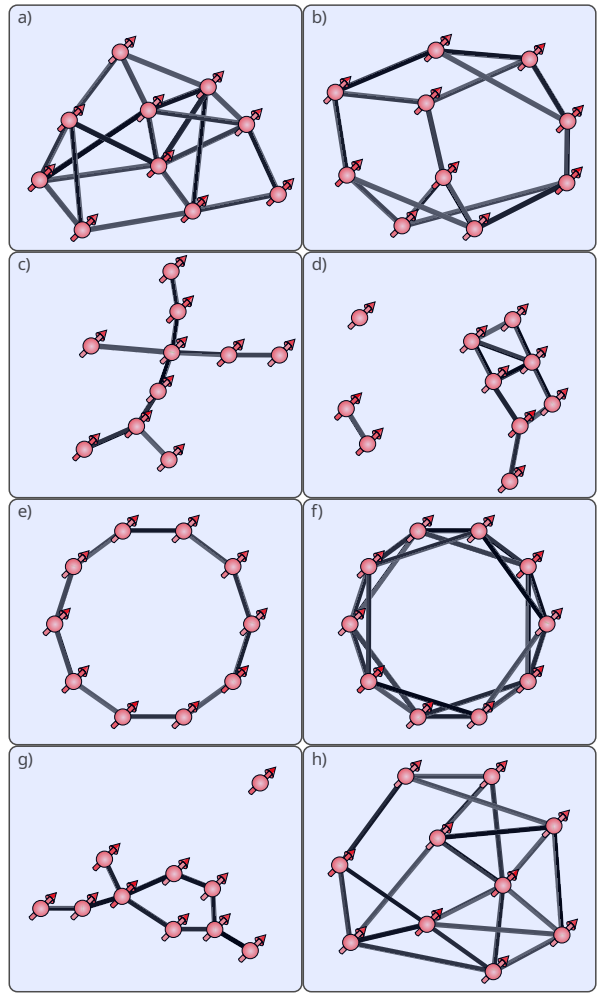}
\caption{Illustrations of randomly generated graphs with $|\mathcal{V}|=10$ vertices. The edges are given without their weight. The graphs were generated using the model: \textbf{a)} Erdős–Rényi with $G^\text{ER}(\mathcal{V}, |\mathcal{E}|=2|\mathcal{V}|)$, \textbf{b)} 3-regular with $G^\text{3-reg}(\mathcal{V}, |\mathcal{E}|=\frac{3|\mathcal{V}|}{2})$ \textbf{c)} Barabasi-Albert with $G^\text{B}(\mathcal{V}, |\mathcal{E}|=|\mathcal{V}|)$ \textbf{d)} Erdős–Rényi with $G^\text{ER}(\mathcal{V}, |\mathcal{E}|=|\mathcal{V}|)$, \textbf{e)} Regular ring lattice with $G^\text{Ring}(\mathcal{V}, |\mathcal{E}|=|\mathcal{V}|)$ \textbf{f)} Regular ring lattice with $G^\text{Ring}(\mathcal{V}, |\mathcal{E}|=2|\mathcal{V}|)$, \textbf{g)} Watts-Strogatz with $G^\text{W}(\mathcal{V}, |\mathcal{E}|=|\mathcal{V}|)$, \textbf{h)} Watts-Strogatz with $G^\text{W}(\mathcal{V}, |\mathcal{E}|=2|\mathcal{V}|)$}
\label{fig:multiple_graphs}
\end{figure}

In the simulations of \cref{sec:simulations} we use various generated random graphs. At the start, we generate the different structures of these undirected random graphs as described below, see \cref{fig:multiple_graphs} for a visual representation. Afterward, all edges receive a uniformly chosen random weight, $w_{i,j}\in[-1,1]$.\\
\vphantom{y}\\
\textbf{Regular graph} $G^\text{k-reg}(\mathcal{V}, \mathcal{E})$: A graph is called $k$-regular if the property $\deg(v_i)=k$ is satisfied for all $v_i$. In \cref{sec:simulations} we use random $3$-regular graphs. Note that, due to the usage of a fixed degree sequence method, the sampling is not uniform.\\
\textbf{Erdős–Rényi model} $G^\text{ER}(\mathcal{V}, \mathcal{E})$: The Erdős–Rényi model \cite{Erdos} describes one of two methods to generate random graphs. The method used in our simulation generates a graph $G^\text{ER}(\mathcal{V},|\mathcal{E}|)$ via the input of the node set $\mathcal{V}$ and the cardinality of the edge set. A graph is picked from the set of all graphs having $|\mathcal{V}|$ nodes and $|\mathcal{E}|$ edges with uniform probability excluding self-loops. Erdős–Rényi graphs are inherently disordered and the probability of finding an edge between a pair of vertices is the same for all pairs, i.e. all vertices have the same average degree $\left<\deg(v_i)\right>=\frac{2|\mathcal{E}|}{|\mathcal{V}|}$. Thus, such graphs tend not to have a strong community structure. In the simulations, we use three different setups. That is, $|\mathcal{E}|=|\mathcal{V}|,\left\lfloor \frac{3}{2}|\mathcal{V}|\right\rfloor, 2|\mathcal{V}|$.\\
\textbf{Barabási–Albert} $G^\text{B}(\mathcal{V}, \mathcal{E})$: The Barabási–Albert model \cite{barabasi} uses the principles of expanding networks and preferential attachment to generate random scale-free networks. Starting with a nucleus of $m_0$ vertices and a positive integer $m$, each time step introduces a new vertex, which is connected to $m$ already existing vertices. The probability of connecting the new vertex to an old one $v_i$ is proportional to its degree $p_i=\nicefrac{k_i}{\sum_j^n k_j}$. With $n$ being the current network size. The most prominent property is the degree distribution, which follows a scale-free power law $P(k)\sim k^{-3}$. In \cref{sec:simulations} we use $m=1$ and $m=2$. \\
\textbf{Regular ring lattice} $G^\text{Ring}(\mathcal{V}, \mathcal{E})$: A regular ring lattice is a special case of a regular lattice, in which the vertices are arranged in a circle and each vertex is connected to its $k=\deg(v_i)\,\,\forall i$ nearest neighbors. Thus, it follows that the degree $k$ must be even. In \cref{sec:simulations} we use $k=2$ and $k=4$.\\
\textbf{Watts-Strogatz} $G^\text{W}(\mathcal{V}, \mathcal{E})$: The Watts-Strogatz \cite{Watts1998} model is an algorithm to interpolate between a ring lattice and a random graph. Given a number of vertices $n$, an even integer average degree $\left<\deg(v_i)\right>=k$, and a rewiring probability $p$, the algorithm starts from a ring lattice, where each vertex is connected to the $k$ nearest neighbors, that is, $\frac k 2$ on each side. Then, beginning with one vertex, the edge connecting to its clockwise nearest neighbor is rewired with probability $p$, that is, the edge $(v_i,v_{i+1})$ is replaced by $(v_i,v_j)$. The vertex $v_j$ is chosen uniformly at random, with self-loops and duplicate edges not allowed. This process is repeated in a clockwise fashion for every vertex on the ring until one circulation is done, followed by a rewiring of next nearest neighbor edges. After a total of $\frac{k}{2}$ circulations, all edges were considered once. The main properties are a small average path length, defined by $l_G:=\frac{1}{n(n-1)}\sum_{i\neq j}d(v_i,v_j) $, where $d(v_i,v_j)$ denotes the shortest distance between the vertices, and a large average local clustering coefficient. For vertex $v_i$, the local clustering coefficient is the fraction of edges connecting two vertices in the neighborhood of $v_i$ over the number of possible edges. In our simulations, we chose $p=0.3$ and two versions with $\left<\deg(v_i)\right>=2$ and $\left<\deg(v_i)\right>=4$, respectively.    

\section{Comments on \texorpdfstring{$\eta$}{eta}}\label{sec:comments_on_eta}
\begin{figure*}[!ht]
\begin{minipage}[c]{0.49\textwidth}
\raggedright
\textbf{a)}\\
\includegraphics[scale=0.56]{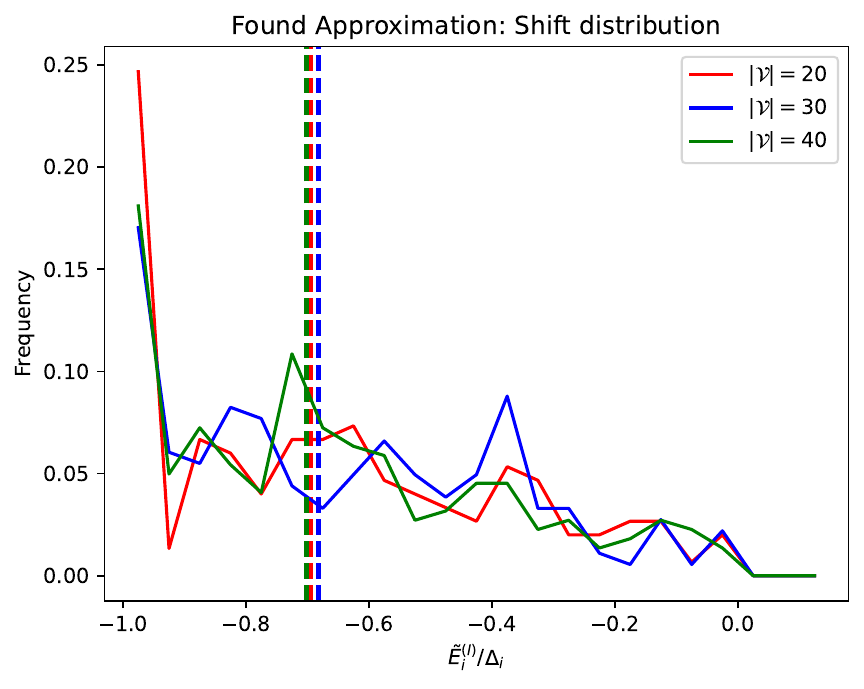}
\end{minipage}
\begin{minipage}[c]{0.49\textwidth}
\raggedright
\textbf{b)}\\
\includegraphics[scale=0.56]{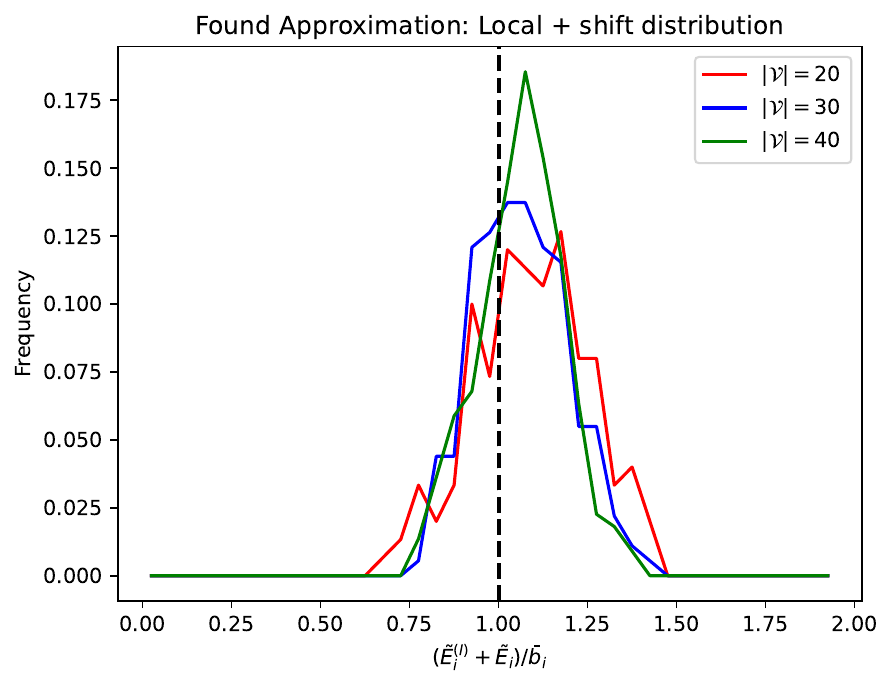}
\end{minipage}
\caption{The found approximations to the global ground state $\ket{\tilde{\psi}_0}$ of the weighted random 3-regular instances of \cref{fig:eta_variation}, when using $\eta=0.5$, were used to create the histograms: \textbf{a)} The interaction energy $\tilde{E}_i^{(I)}$ over the bound of the interaction energy $\Delta_i$  (for communities in the first iteration of the algorithm). The vertical lines indicate the median; \textbf{b)} The interaction energy combined with the local energy of the found approximation over the bound for $\bar{\mathscr{F}}_i$ states.}
\label{fig:shift_distro_plots}
\end{figure*}

In this section, we discuss the observation that reducing the energy range by a factor $\eta < 1$ has only a very small effect on the optimization ratio down to a certain value for $\eta$. This can be observed in \cref{sec:simulations}, where $\eta=0.5$ retains $99.9\%$ of the approximation ratio on weighted random 3-regular graphs with 40 vertices. 

Let us focus on one community $\Com_i$ and again denote the states of $\Com_i$ that have been kept by $\mathscr{F}_i$ and the states that would have been kept with the full range, but are excluded by $\eta < 1$, as $\bar{\mathscr{F}}_i$. For simplicity, we assume that the state of the environment is fixed with energy $E_\mathscr{E}$. Then the optimal energy we get with the reduced range is given by 
\begin{equation}
    E(\eta) = \min_{\ket{\psi^\mu_i}\in\mathscr{F}_i}(E^\mu_i+E^{(I),\mu}_i+E_\mathscr{E})
\end{equation}
where $E^{(I),\mu}_i= \bra{\psi_i^\mu,\psi_\mathscr{E}}H_i^{(I)}\ket{\psi_i^\mu,\psi_\mathscr{E}}$, and the energy we get when only considering the neglected part is given by
\begin{equation}
    \bar{E}(\eta) = \min_{\ket{\psi^\mu_i}\in\bar{\mathscr{F}}_i}(E^\mu_i+ E^{(I),\mu}_i+E_\mathscr{E}) .
\end{equation}
The energy we miss out on, when reducing $\eta$ is zero when $E(\eta)<\bar{E}(\eta)$ and their difference
\begin{align}
    E(\eta)-\bar{E}(\eta)=&\min_{\ket{\psi^\mu_i}\in\mathscr{F}_i}(E^\mu_i+E^{(I),\mu}_i) - \nonumber\\
     &\min_{\ket{\psi^\mu_i}\in\bar{\mathscr{F}}_i}(E^\mu_i+E^{(I),\mu}_i)
\end{align}
otherwise.
The first effect that could lead to a good approximation ratio $\alpha$ is that one of the states in $\mathscr{F}_i$ is shifted by a large enough amount s.t. $E(\eta)$ is lower than or equal to the lowest possible value of $\bar{E}(\eta)$, i.e.
\begin{equation}
    E(\eta) \leq \bar{b}_i +E_\mathscr{E} .
\end{equation}
where we have introduced the notation $\bar{b}_i:= E^0_i+ (\eta-1) \Delta_i$ for the lower bound.

The second effect that could lead to a good $\alpha$ is that $\bar{E}(\eta)$ does not get small enough (via the interaction shift) to contribute to the ground state. Indeed, consider the ground state $\ket{\psi^0_i}$ that is shifted by some negative expected value $\mathds{E}[E^{(I)}]$, then to merely beat the local ground state it is required that there exists an interaction with the environment such that $\bar{E}^{(I)}_i<\eta \Delta_i+\mathds{E}[E^{(I)}]$. In addition, states in $\bar{\mathscr{F}}_i$ that do not have a local energy at the lower boundary of the distribution would need an even greater negative shift. For distributions that decay very fast, this is unlikely. Also, note that the energies of the states in $\mathscr{F}_i$ and $\bar{\mathscr{F}}_i$ are correlated, and analyzing the distribution of the interaction Hamiltonian gives limited insight. 

To investigate these effects, we analyzed the instances of random 3-regular graphs that were also used in \cref{fig:eta_variation}. For the algorithm's output states $\tilde{\ket{\psi_0}}$ for $\eta=0.5$, which yielded good approximations, we checked the strength of the interaction shifts $\tilde{E}_i^{(I)}$ for the communities in the first iteration of the algorithm and compared them to their bound $\Delta_i$, see \cref{fig:shift_distro_plots} a).
We see that large relative interaction shifts are dominant but not exclusive. In addition, the shift distribution does not change significantly when increasing the total size of the system from $|\mathcal{V}|=20$ to $|\mathcal{V}|=40$.

In \cref{fig:shift_distro_plots} b) we combined the interaction energy $\tilde{E}_i^{(I)}$ with the local energy $\tilde{E}_i$ and compared it with the lower bound $\bar{b}_i$ for the cut states. Here, values larger than 1 indicate that the first effect guarantees that no loss of approximation ratio/energy is possible, for a given environment state, which occurs in the majority of all cases. Values lower than 1 indicate the possible existence of a state in $\bar{\mathscr{F}}_i$ that has a lower overall energy. However, from the result in \cref{fig:eta_variation} we know that this is very unlikely. After further investigation, we found that some instances have $\bar{\mathscr{F}}_i=\varnothing$, but for the majority of communities in this category, the states in $\bar{\mathscr{F}}_i$ simply do not have a strong enough interaction. Whether this is due to a low probability in the interaction energy distribution or to correlations is hard to say from this analysis. 

We conclude that strong interaction shifts for low lying states are the dominant factor for the resulting good optimization ratios, but see that they are not the only reason. Due to the complex nature of the problem, it is hard to make statements about the scaling of this effect. The results of \cref{sec:simulations} at least show that the effect is not exclusive to weighted random 3-regular graphs for the system sizes considered.

\bibliographystyle{apsrev4-2}
\bibliography{bibliography}

\end{document}